\documentclass[preprint]{aastex}

\bibpunct[]{(}{)}{;}{a}{}{,}

\newcommand{\etal}{et al.\ }
\newcommand{\kms}{\, {\rm km\, s}^{-1}}

\newcommand{\lya}{Ly$\alpha$ }
\newcommand{\bF}{\bar{F}}

\newcommand{\heii}{He II }
\newcommand{\hi}{H I}
\newcommand{\Db}{\Delta_b}
\newcommand{\gmo}{\gamma-1}

\slugcomment{Submitted to ApJ}

\shorttitle{Observed Flux Statistics in the \lya Forest}
\shortauthors{McDonald \etal}

\begin{document}

\title{The Observed Probability Distribution Function, Power Spectrum, 
and Correlation Function of the Transmitted Flux in the \lya Forest
\altaffilmark{1}}

\author{Patrick McDonald,\altaffilmark{2}
Jordi Miralda-Escud\'e,\altaffilmark{2,3}
Michael Rauch,\altaffilmark{4} 
Wallace L. W. Sargent,\altaffilmark{5} 
Tom A. Barlow,\altaffilmark{5}
Renyue Cen,\altaffilmark{6}
and Jeremiah P. Ostriker\altaffilmark{6}}

\altaffiltext{1}{The observations were made at the W.M. Keck 
Observatory, which is 
operated as a scientific partnership between the California Institute 
of 
Technology and the University of California; it was made possible by
the generous support of the W.M. Keck Foundation. }
\altaffiltext{2}{Department of Physics and Astronomy, University of 
Pennsylvania, Philadelphia, PA 19104; 
pmcdonal,jordi@llull.physics.upenn.edu}
\altaffiltext{3}{Alfred P. Sloan Fellow}
\altaffiltext{4}{ESO, Karl-Schwarzschild-Str. 2, 85748 Garching, 
Germany}
\altaffiltext{5}{Astronomy Department, California Institute of 
Technology, Pasadena, CA 91125}
\altaffiltext{6}{Princeton University Observatory, Peyton Hall, 
Princeton, NJ 08544}

\begin{abstract}

A sample of eight quasars observed at high resolution and signal-to-noise is
used to determine the probability distribution function (PDF), the power
spectrum, and the correlation function of the transmitted flux in the \lya
forest, in three redshift bins centered at $z=2.41$, $3.00$, and $3.89$. All
the results are presented in tabular form, with full error covariance matrices
to allow for comparisons with any numerical simulations and with other data
sets. The observations are compared with a numerical simulation of the \lya
forest of a $\Lambda$CDM model with $\Omega=0.4$, known to agree with other
large-scale structure observational constraints.
There is excellent agreement for the PDF, if the
mean transmitted flux is adjusted to match the observations. A small difference
between the observed and predicted PDF is found at high fluxes and low
redshift, which may be due to the uncertain effects of fitting the spectral
continuum. Using the numerical simulation, we show how the flux power spectrum
can be used to recover the initial power spectrum of density fluctuations. From
our sample of eight quasars, we measure the amplitude of the mass power
spectrum to correspond to a linear variance per unit $\ln k$ of $\Delta^2_\rho
(k)=0.72\pm 0.09$ at $k = 0.04 (\kms)^{-1}$ and $z=3$, and the slope of the
power spectrum near the same $k$ to be $n_p=-2.55 \pm 0.10$ (statistical error
bars). The results are statistically consistent with \citeauthor{cwp99},
although our value for the rms fluctuation is lower by a factor $0.75$.
For the $\Lambda$CDM model we use, the implied primordial slope is
$n=0.93\pm 0.10$, and the normalization is
$\sigma_8=0.68 + 1.16(0.95-n) \pm 0.04$.

\end{abstract}

\keywords{
cosmology: observations---intergalactic medium---large-scale structure
of universe---quasars: absorption lines
}

\section{INTRODUCTION}

  The study of the \lya forest has been making a transition toward
unification with other methods of investigating the large-scale 
structure of the universe. 
Several analytical models had proposed that the structures
formed at high redshift through gravitational collapse on a range of scales
and gas densities could produce the \lya forest absorption lines 
\citep*{bs65,a72,r86,bss88,m90,b93,mr93}. Detailed 
hydrodynamical
simulations of the evolution of structure in a photoionized intergalactic
medium (hereafter, IGM) in cold dark matter models have shown that the
basic properties of the \lya absorption spectra 
\citep[see][for a review]{r98} can indeed be understood by the 
evolving network of sheets,
filaments, and halos characteristic of gravitational dynamics in 
cosmology
\citep*[e.g.,][]{cmo94,zan95,hkw96,mco96,wb96,zan97,tle98}.
Progress with theoretical modeling of the intergalactic medium 
occurred at the same time as the first high resolution and signal-to-noise 
Ly$\alpha$ forest spectra from the Keck telescope's HIRES instrument 
\citep{vab94} became available, and as the large transverse sizes
\citep*{b94,d94} and the small-scale smoothness \citep*{s92,s95} of the
absorbers were discovered, which eliminated many of the alternative models.

 As the results of these simulations and the effects of numerical
resolution, box size and thermal evolution of the IGM are better
understood, one can start measuring parameters of the theory of large-scale
structure from the observations of the \lya forest. Thus, the distribution
and the mean of the transmitted flux constrain the density-temperature
distribution of the ionized gas, and determine a parameter depending mainly
on the baryon density and the intensity of the ionizing background 
\citep{rms97,wmh97}. The power spectrum of the \lya forest is
closely related to the power spectrum of the mass fluctuations, allowing
a measurement of the amplitude of these fluctuations at high redshift
which gives important constraints for large-scale structure models 
\citep*{cwk98,cwp99,wch99,chd99}. The Doppler parameter
distribution of the absorption lines can be used to measure the
mean temperature at different densities \citep*{stl99,rgs99,bm99}, 
which depends on the ionization history of the IGM
\citep{hg97}.

  A lot of these observational results are based on a direct measurement of
the statistical properties of the transmitted flux in the \lya forest,
which is a one-dimensional random field depending on the density,
temperature and velocity of the gas at every point along the line of sight.
While a large body of observational data has been published following the
more traditional method of fitting the absorption as being due to discrete
absorbers with Voigt profiles, the work using flux statistics was generally
aimed at determining specific cosmological information and less focused
on
a general presentation of the observational results. In this paper, we
present the first detailed tabulation of the probability distribution
function, power spectrum, and correlation function of the transmitted flux
using the highest quality spectra currently available, and paying special
attention to the calculation of error bars. The primary focus of this paper
is on a clear presentation of the observational results, in a form that is
useful for comparisons with future cosmological simulations and models, as
well as other observations. With this aim, the statistics that we use are
intended to contain the minimum amount of complexity and theoretical
prejudice. We will also compare our observational results to the numerical
simulation ``L10'' in \citet{mco96}, and discuss the
cosmological implications. We demonstrate how our results can be used to
measure the primordial power spectrum by a method similar to 
\citet{cwp99}. Our data has much better resolution, 
$\sim 0.1$ \AA, so we can push
the measurement of $P(k)$ down to wavelengths $\lambda \sim 1 h^{-1}$ Mpc. 

In \S 2 we describe the data and the simulation that we use.
In \S 3 we present the results for the mean and variance of the
transmitted flux and discuss their implications for cosmological 
parameters.
In \S 4 we present the probability distribution function
of the transmitted flux and compare it to that of the
numerical simulation.  
In \S 5 we present the 
results for the power spectrum of the transmitted flux, and compare them
to the simulation. We then discuss the implications for the
primordial power spectrum of the universe, from the results on scales that
are large enough to make the \lya forest fluctuations be related to
quasi-linear density fluctuations.
In \S 6 we present the correlation function of the
transmitted flux.
Finally, the discussion and conclusions are given in 
\S\ref{discussion}.
Appendix B describes the computation of the error bars.
All the results in this paper are available in the website
\url{http://www.physics.upenn.edu/$\tilde{~}$jordi/lya}. In 
addition 
to all the
Tables included here, the website contains also the full error
covariance matrices and other details that are too extensive to be
published here.

\section{THE OBSERVATIONAL DATA AND THE SIMULATION}

In this section we describe the observational data set and the 
\lya forest simulation that we use.

\subsection{Description of the Observations}

We use a set of eight quasars with spectra that are fully resolved
and have high signal-to-noise ratio. 
Seven of our quasars (Q2343+123, Q1442+293, Q1107+485, Q1425+604,
Q1422+230, Q0000-262, and Q2237-061) are
the same as in \citet{rms97}, but we add 
KP 77: 1623+2653, one of the triplet of quasars described in 
\citet{cf98}.  
The pixel noise is typically less than 5\% of the continuum flux level,
and frequently as low as 1\%.
The velocity resolution is 6.6 $\kms$ (FWHM) and the spectra are 
binned in 0.04 \AA~ pixels.

Continua were fitted to the HIRES spectra using the IRAF Continuum
task.  Spline3 or chebycheff polynomials were used. The order of the
polynomials and the number of fitting regions strongly depended on the
signal-to-noise in the spectra and the redshift. As a rule of thumb,
the higher the S/N ratio and the weaker the absorption, the more fitting
regions were used and the higher the order of a polynomial was taken
(amounting to several tens of degrees of freedom for a typical case of
S/N$\sim$ 30, $z = 2.5$ Ly$\alpha$ forest spectrum).
The regions between Ly$\alpha$ and Ly$\beta$ were cut into 2 to 4 pieces
for fitting, which were re-joined after fitting to form a long spectrum.

The continua were fitted to regions of the Lyman $\alpha$ forest deemed
'free of absorption lines' (as judged by eye). At best this is a
problematic definition, as the perception of a stretch of spectrum as
being free of absorption depends on the signal-to-noise ratio of
the data, the average line density (and thus the redshift), the mean
background absorption (i.e., the possibility of a constant absorption
trough on top of individual absorption lines), and the spectral
resolution.  The last point does not pose a problem for the current
data as absorption features with widths down to 6.6 kms$^{-1}$ are
resolved, and  typical Lyman $\alpha$ line widths are much larger than that
\citep[e.g.,][]{rcc92}. The possibility of unrecognized large-scale
absorption features in the absorption is a more serious concern.
Determining the position and
extent of line-free spectral regions by eye tends to mistake large
shallow flux depressions as parts of the unabsorbed continuum, so the
continuum would be systematically placed low, leading to an
underestimate of the absorbed optical depth.  Similar problems arise
for the high redshift (z$>$ 4) Ly$\alpha$ forest where more than half
of the flux is absorbed and the spectrum 'between the absorption lines'
rarely recovers enough to reach the probable continuum level of the
QSO.  In such cases only a stiff polynomial (typically with 3 to 5 degrees
of freedom for the region from Ly$\alpha$ to Ly$\beta$) can be used.
This lack of information caused by the sparsely sampled continuum can
lead to local uncertainties in the continuum exceeding 5\% , as can be
shown by comparison with more easily flux-normalizeable, low resolution
single order spectra of the same QSO. The S/N ratio, finally, may also
lead to biases, in that, at higher redshifts, the apparently unabsorbed
continuum portions between the absorption lines shrink in size and
noise. As the final selection of fitting regions is done by eye, the
shortness of the continuum portions could easily eliminate true
continuum regions or introduce spurious ones in the sample of supposed
continuum data points. This last effect does not produce global errors
in the continuum level, but it can introduce local fluctuations.

  These uncertainties in the continuum fitting can affect the results
reported in this paper on the flux distribution function, the power
spectrum and the correlation function, and should be considered as a
source of possible systematic errors in addition to quoted statistical
errors of the results. This underscores the need to obtain spectra in
the future with a good flux calibration, so that the continuum can be
fitted with many fewer free parameters under the assumption of a
smooth underlying continuum spectrum of the observed quasar \cite{prs93}.

\begin{figure}
\plotone{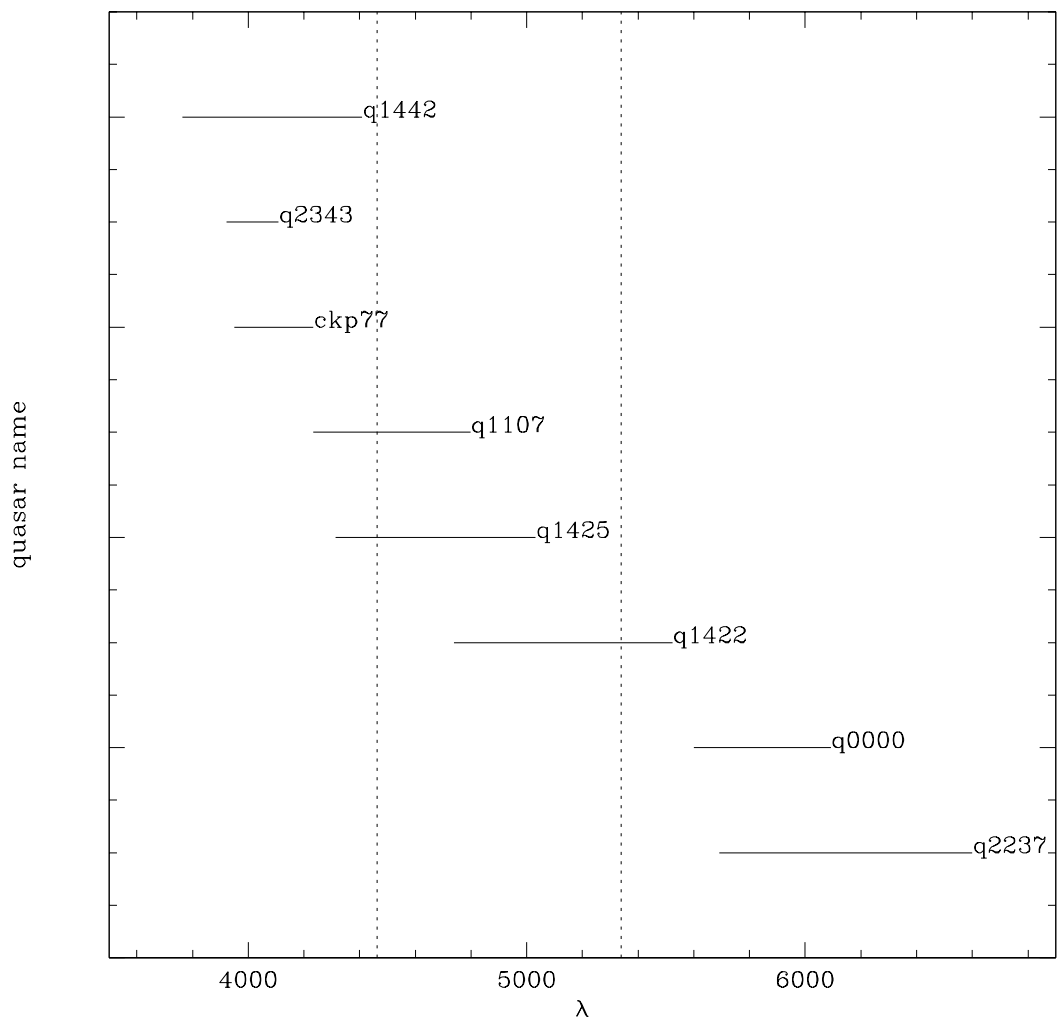}
\caption{The useful redshift range covered by each quasar.
The vertical axis is meaningless.  The vertical dotted lines 
at $z=2.67$ and $z=3.39$ separate
the data into the three redshift bins that we will use.
}
\label{quasars}
\end{figure}
Figure \ref{quasars} shows the redshift
range covered by the useful part of each spectrum.
The wavelength regions of the spectra of these eight quasars that were 
included in the present analysis were selected as follows: first, only 
the regions between the \lya and Ly$\beta$ emission wavelengths were 
included, excluding also an interval within 5 $h^{-1}$ Mpc of the 
quasar to avoid the proximity effect. 
Approximately half of the
KP 77 spectrum has signal-to-noise ratio lower than any of the 
\citet{rms97} data so this region was eliminated.  
Finally, damped \lya absorption lines and narrow lines known or
suspected to be metal lines were removed, by eliminating small intervals
of the spectra containing these absorption lines from the spectra.
The detailed list of wavelength intervals eliminated in each quasar
is available on the website mentioned above.
The data preparation is described in more detail in 
\citet{rms97} and \citet{bs97}.

\subsection{Calculation of Errors}

We present in this paper the results on the flux distribution function
\citep[in greater detail than in][]{rms97}, as well as the power
spectrum and correlation function of the transmitted flux. All the
observational results are given with error bars obtained using the
bootstrap method \citep{ptv92}, including the full covariance matrices
for all the results except the power spectrum.
The bootstrap procedure consists of 
dividing the data into $N$ segments and generating modified 
realizations of a statistic by randomly selecting $N$ of the segments 
(with replacement). The dispersion in the bootstrap realizations
approximates the error on the observed statistic. The
method for computing the error bars is described in detail in
Appendix B.
We present only the diagonal elements of the error matrices in the
tables here. The full covariance matrices can be obtained from the website
mentioned above. Our 
intention is to allow detailed comparisons of these results with
cosmological simulations and with other observational results.

\subsection{Description of the Simulation}

  We compare the observations to the output of the Eulerian 
hydrodynamical simulation described in 
\citet{mco96} (referred to as L10 in that paper).
The cosmological model used is
$\Omega_0=0.4$, $\Omega_\Lambda=0.6$, $h=0.65$, $\sigma_8=0.79$, and
a large-scale primordial power spectrum slope $n=0.95$. This model is in
agreement with the large set of observations of large-scale structure
currently available \cite[e.g.,][]{wcos99}. The box size 
of the simulation is $10 h^{-1}$ Mpc, and it contains $288^3$ cells.
\lya spectra are computed for a large number of lines of sight along the
box axes. There is one free parameter that we can vary when computing
the spectra, the normalization of the optical depth, which we adjust to
reproduce the observed transmitted flux, as we show in \S 3.
Renormalizing the optical depth is equivalent to modifying the intensity
of the ionizing background, as long as the effect of collisional
ionization and the change in the gas temperature caused by the different
heating rate can be neglected
\citep[see][for a test that these effects are in fact
negligible; notice that collisional ionization is actually important
in high temperature gas in the simulations, but this gas is always at
high density and produces saturated absorption in \lya]{tle98}. 
The optical 
depth is then mapped to transmitted flux using $F=\exp(-\tau)$.

  These theoretical spectra are then modified to account for the
continuum fitting, resolution and noise in the observations. Of these
three effects, the most difficult to reproduce is the operation of
continuum fitting, where several points along the spectrum of the
quasars that appear not to have any obvious absorption are selected to
indicate the quasar continuum. This operation is inevitable if the
intrinsic spectrum of the quasar is unknown.
Unfortunately, there is not really a good
way to correct for it because the spectra obtained from simulations
obviously have the period of the simulated box, which is comparable to
the typical distance between successive continuum fitting points. 

  Following \citet{rms97}, we estimate the effects of continuum 
fitting by defining the maximum transmitted flux along any line of
sight in the simulation (parallel to one of the three axes) to be
the continuum flux, $F_c$, and dividing the flux in all other pixels 
by $F_c$.
We convolve with the instrumental resolution of $6.6 \kms$, and
we add Gaussian noise to each cell with the dispersion $n(F)$, which is
given below
in Table \ref{pdftab}.
The dispersion $n(F)$ is computed as the mean noise in all pixels in 
the observed spectra for every flux bin.
Notice that the noise increases with the flux value because
there is a constant background noise plus the Poisson noise of the
photon counts in each pixel.
We also map the 288 cells along an axis in the simulation
onto 512 pixels in the spectra, to facilitate the computation of Fourier
transforms.
The noise in Table \ref{pdftab} is for the pixel width in the
observational data, $0.04$ \AA.
When we compute the power
spectrum and correlation function this noise level is multiplied by 
$(0.04~{\rm \AA} / \Delta \lambda_{sim})^{1/2}$ where 
$\Delta \lambda_{sim}$
is the wavelength extent of the pixels in the simulated spectra.

  In \citet{rms97}, an additional correction to the observed
spectra was applied to correct for the redshift evolution within a
redshift bin before a
comparison is made with a simulation at a given redshift $z_i$. This
consisted of multiplying the optical depth in pixels within a redshift
bin around $z_i$ by the factor $[(1+z)/(1+z_i)]^{4.5}$, so that
all the observed pixels are corrected to a redshift $z_i$ according to
a law that approximately fits the observed evolution. We do not apply
this correction here. The correction is small for the redshift bins
we shall use; nevertheless, it must be born in mind that our
observational results for the flux distribution and power spectrum are
averages over each of our redshift bins. The reason we do not include
this correction is that it changes the statistical distribution of the
pixel noise in a way that biases some of the results. Not including the
correction also makes it easier to compare to any simulation in an
accurate way: after the continuum fitting, resolution and noise are
included in simulated spectra, one can average the quantity being
compared over the same redshift bins used here.

\section{THE MEAN AND VARIANCE OF THE TRANSMITTED FLUX}

In this section we present results for the mean and variance of the
transmitted flux, with error bars calculated using the
resampling method described in Appendix B. We then revisit the value
of $\Omega_b$ derived in \citet{rms97}.

The mean transmitted flux $\bar{F}$ is computed by averaging over all
the pixels in a certain data subset (either a given quasar or a 
redshift
bin).
\begin{figure}
\plotone{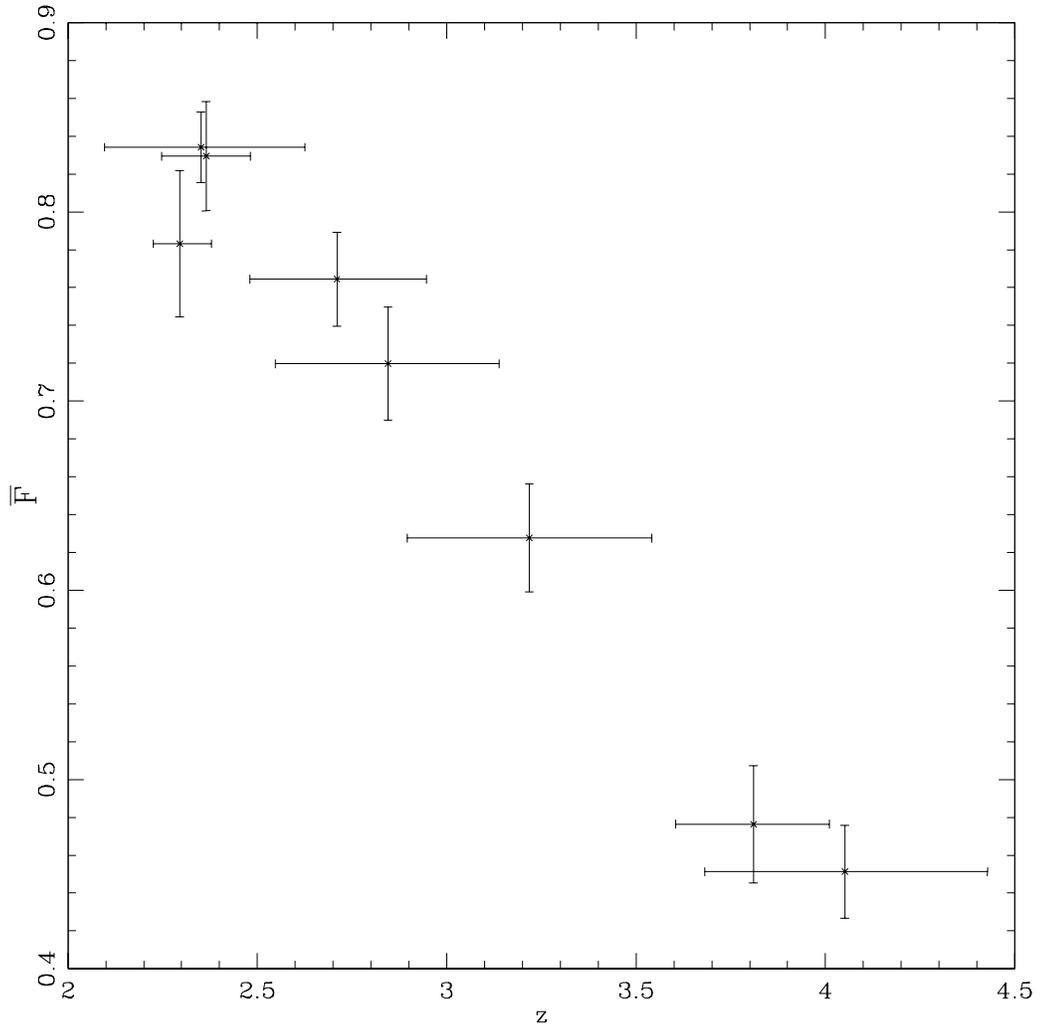}
\caption{Mean transmitted flux, $\bar{F}=\left<F\right>$, for each 
quasar in the sample.  The 
error bars in the $z$ direction show the redshift range covered by
each spectrum while the point shows the mean redshift. }
\label{meanFs}
\end{figure}
Figure \ref{meanFs} shows the mean transmitted flux with error bars
for each quasar.  
\begin{figure}
\plotone{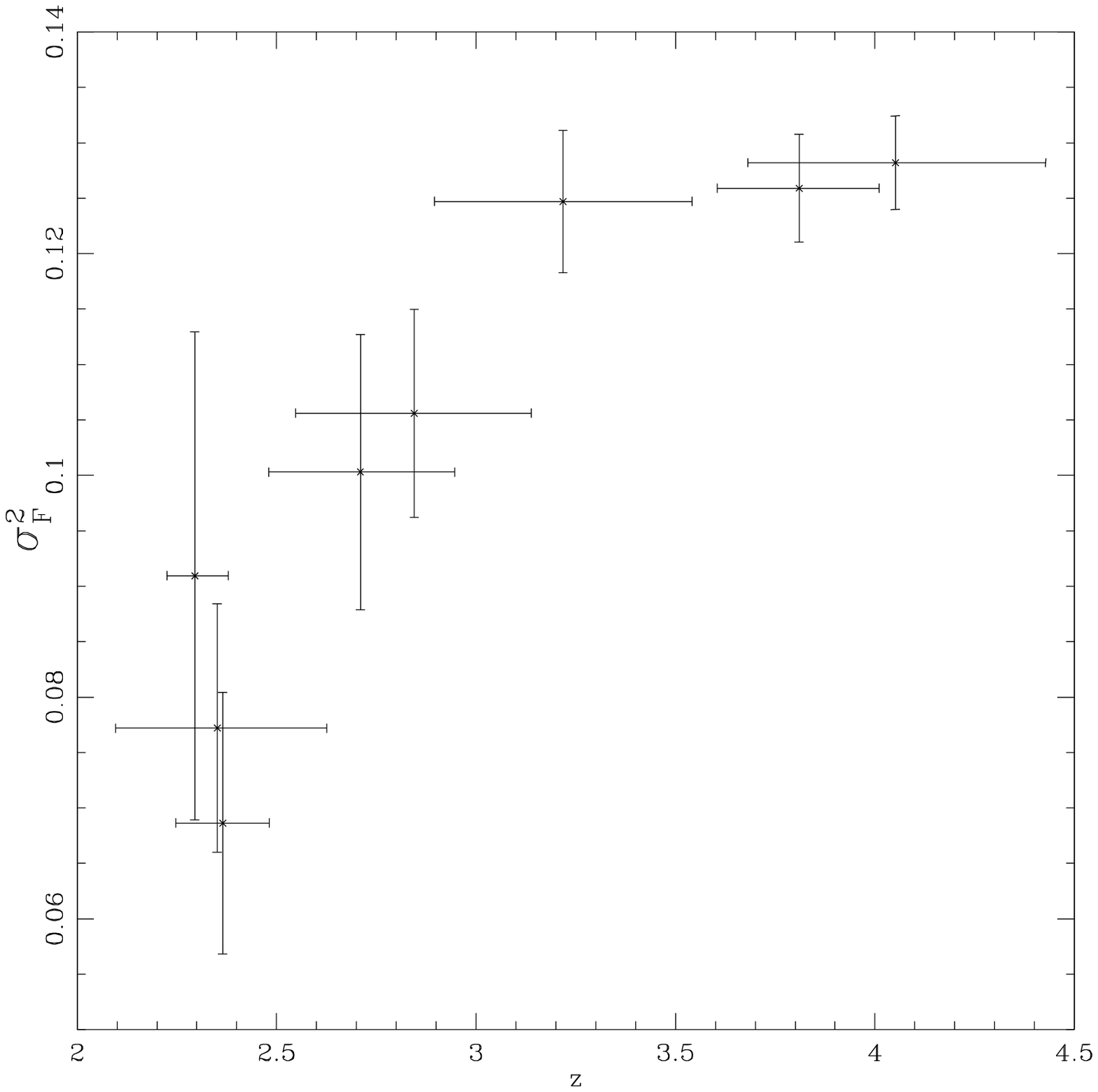}
\caption{Variance of the transmitted flux, 
$\sigma_F^2=\left<(F-\bar{F})^2\right>-\sigma_{noise}^2$, for each 
quasar in the sample.  The 
error bars in the $z$ direction show the redshift range covered by
each spectrum while the point shows the mean redshift. 
The variance decreases with decreasing redshift because the mean 
transmitted flux is increasing.}
\label{sigsqs}
\end{figure}
Figure \ref{sigsqs} shows the 
variance of the flux for each quasar, 
$\sigma_F^2=\left<(F-\bar{F})^2\right>-
\sigma_{noise}^2$.  The variance contributed by the pixel noise, 
$\sigma_{noise}^2$, is generally
negligible.  
The pixel noise does not significantly affect the error bars
on the mean and variance of the transmitted flux, which are 
dominated by the fluctuations in the number of absorbers in the
spectra.
We notice that the increase of $\sigma_F^2$ with redshift is due to
the increasing mean flux decrement (the true density fluctuations are
of course decreasing with redshift due to gravitational clustering
evolution). 

\begin{deluxetable}{ccccccc}
\tablecaption{Mean and variance of the transmitted flux. 
\label{meanFtab}}
\tablehead{
\colhead{$z_{min}$} & \colhead{$z_{max}$} & \colhead{$\bar{z}$} & 
\colhead{$\bar{F}$} & \colhead{$\sigma^2_F$} &
\colhead{$\left< \delta \bar{F} \delta \sigma^2_F \right>$} 
& \colhead{pixels}}
\startdata
3.39 & 4.43 & 3.89 & $0.475 \pm 0.021$ & $0.1293 \pm 0.0030$ &
$-1.0\times 10^{-5}$ & 34320 \\
2.67 & 3.39 & 3.00 & $0.684 \pm 0.023$ & $0.1174\pm 0.0056$ &
$-1.1\times 10^{-4}$ & 31897 \\ 
2.09 & 2.67 & 2.41 & $0.818 \pm 0.012$ & $0.0789\pm 0.0068$ &
$-7.7\times 10^{-5}$ & 33791\\ 
\enddata
\tablecomments{The mean flux decrement $\bF$, the flux
variance $\sigma^2_F$, their error correlation, and the total number
of data pixels are listed for each redshift bin ($z_{min}$, $z_{max}$),
with mean redshift $\bar{z}$.}
\end{deluxetable}
The results for $\bF$ and $\sigma_F^2$ are presented in
three redshift bins, centered at $\bar{z}=2.41$, $\bar{z}=3.00$,
and $\bar{z}=3.89$, in Table \ref{meanFtab}.
Each of these bins contains about one third of the data. 
We will use the same redshift bins for presenting observational 
results in the rest of the paper. 
The decrease in fluctuations with increasing $\bF$ accounts for 
the smaller error bar on the mean transmitted flux at $\bar{z}=2.41$.

\subsection{Cosmological Implications of the Mean Flux Decrement
Revisited}

The value of the mean transmitted flux is related to the parameter:
\begin{equation}
\mu = \left(\frac{\Omega_B h^2}{0.0125}\right)^2 \left[\frac{100 \kms 
{\rm Mpc}^{-1}}{H(z)}\right]\left(\frac{1}{\Gamma_{-12}}\right) ~,
\label{mudef}
\end{equation}
where $\Gamma=10^{-12} \, \Gamma_{-12}\, {\rm s}^{-1}$ is the
photoionization rate due to the cosmic ionizing background. The
constant $\mu$ includes the simple dependences on cosmological
parameters that arise from photoionization equilibrium:
if the spatial distribution of the overdensity, temperature and
peculiar velocity of the gas is not altered, the optical depth at
every pixel is proportional to $\mu$ when $\Omega_b$, $H_0$ and
$\Gamma$ are varied.

\begin{deluxetable}{ccccccc}
\tablecolumns{7}
\tablecaption{The optical depth normalization
$\mu\propto (\Omega_b h^2)^2 (H(z)~\Gamma_{-12})^{-1}$
\label{tautable}}
\tablehead{
\colhead{$\bar{z}_{obs}$} & \colhead{$z_{sim}$} & \colhead{$\mu$ (cc)}
& \colhead{$\mu$ (ncc)} &
\colhead{$\Gamma_{-12}$} & \colhead{$\Omega_b~h^2$} 
& \colhead{$T_{0,sim}$}\\
&&&&&&(K) }
\startdata
2.41 & 2 & $1.58\pm 0.20$ & $1.51\pm 0.19$ &
$0.545\pm 0.070$& $0.0257\pm 0.0017$ & 13100\\ 
2.41 & 3 & $1.36\pm 0.17$ & $1.30\pm 0.16$ &
$0.630\pm 0.082$& $0.0239\pm 0.0016$ & 16000\\
3.00 & 3 & $1.66\pm 0.28$ & $1.51\pm 0.23$ &
$0.412\pm 0.068$& $0.0296\pm 0.0024$ & 16000\\
3.00 & 4 & $1.34\pm 0.21$ & $1.20\pm 0.18$ &
$0.514\pm 0.082$& $0.0266\pm 0.0021$ & 14100\\
3.89 & 4 & $1.44\pm 0.17$ & $1.19\pm 0.13$ &
$0.356\pm 0.043$& $0.0319\pm 0.0019$ & 14100\\
\enddata
\tablecomments{Computed for the observational redshift bin $z_{obs}$ by
comparison to the simulation at $z_{sim}$, with (cc) and without (ncc)
the continuum correction. We give $\Gamma_{-12}$ assuming 
$\Omega_b h^2=0.019$, and $\Omega_b h^2$ is given assuming
$\Gamma_{-12}=1$, both including the continuum
correction.  The median temperature at the mean
density for the appropriate simulation output is given as $T_{0,sim}$.}
\end{deluxetable}  
  Table \ref{tautable} presents the value of $\mu$ required for the mean
transmitted flux, predicted directly from our simulation, to match the
observed one, with and without the continuum fitting correction. The
values are given for the same three redshift bins used before (see Table
\ref{meanFtab}). We derive $\mu$ using two simulation outputs for two of
the redshift bins of the observational data. We shall see in \S 5 that
comparing the data to the simulation outputs at different redshifts is
approximately equivalent to varying the amplitude of the power spectrum
in the model, and that the two simulation redshifts used in Table 2
bracket the amplitude that is inferred from our data. In addition to the
amplitude of the power spectrum, the simulation outputs at different
redshifts also differ on the distribution of temperatures. The mean
temperature at overdensity of unity is given in the last column of Table 
\ref{tautable}). The differences in the derived $\mu$ for different
simulation outputs in Table \ref{tautable} are due to both the different
temperatures and the declining fluctuation amplitude with redshift.

  The values of $\mu$ we find are very similar to those in
\citet{rms97}; the slight differences we find when comparing to the same
simulation output are due to the inclusion of a new quasar in our
sample, the different choice of redshift intervals, and differences in
the method of analysis (such as not including the redshift correction to
the center of each redshift bin). We give also in Table \ref{tautable}
the statistical error bars due to the sample variance.
In addition, we show the value of $\Omega_b h^2$ when $\Gamma_{-12}=1$,
and the value of $\Gamma_{-12}$ when $\Omega_b h^2=0.019$, fixing
$H(z)$ to the model $h=0.65$, $\Omega_0=0.4$, $\Lambda_0=0.6$. We
recall here that the observed abundance of quasars provides a lower
limit $\Gamma_{-12} \gtrsim 1$ at $z=3$, and therefore a corresponding
lower limit to $\Omega_b h^2$, which is a bit high compared to the
value derived from nucleosynthesis and the deuterium abundance of
\citet{bt98} \citep[see][]{rms97,wmh97}.

  In addition to the $\mu$ parameter, how should the predicted mean
transmitted flux depend on the specific large-scale structure model that
is assumed? As we shall discuss later in \S \ref{discussion}, the
dominant dependence should be on the gas temperature-density relation,
and the density fluctuation amplitude at the Jeans scale
\citep*[see, e.g.,] {hg97,nh99}. Rescaling the gas
temperature $T_0$ at a fixed overdensity will result in a variation
$\mu \propto T_0^{0.7}$ for a fixed mean transmitted flux, owing to
the variation of the recombination coefficient with temperature. 
Changing the temperature will also change the distribution
of absorption, for a fixed real space distribution of gas, by
altering the thermal broadening. In general, increasing the 
thermal broadening will increase the amount of absorption and so 
decrease $\mu$ for a fixed mean flux decrement. The
full dependence on the gas temperature (which is affected by radiative
cooling, shock heating, etc.) is more complicated; however, these
two effects incorporate most of the temperature dependence.

  The fluctuation amplitude at the Jeans
scale in the simulation we use is probably close to the correct value 
in our Universe given the good agreement we find later (\S 4 and 5) in
the transmitted flux distribution function and the power spectrum. 
This limits the model uncertainty in the value of
$\mu$ derived from the simulation we use. However, the
dependence on the density distribution is strong because 
the density of neutral gas is proportional to the square of the 
baryon density. 

  The different values of $\mu$ derived from different redshift
outputs give an idea of the importance of the model dependence on the
gas temperature and the power spectrum amplitude. A more detailed
analysis using simulations where these parameters are varied will be
necessary to assess the errors due to the theoretical model
uncertainty on the value of $\mu$.

\section{THE PROBABILITY DISTRIBUTION OF THE TRANSMITTED FLUX}

\begin{deluxetable}{ccccccc}
\tablecaption{The observed probability distribution of the 
transmitted flux.
\label{pdftab}}
\tablehead{
\colhead{$F$} & \colhead{$P(F)$ $(z=3.89)$} & \colhead{$n(F)$}& 
\colhead{$P(F)$ $(z=3.00)$} & \colhead{$n(F)$} & \colhead{$P(F)$
$(z=2.41)$} & \colhead{$n(F)$} }
\startdata
0.00 & $3.618\pm0.288$ & 0.024  &$2.032\pm0.229$ & 0.0064 &$0.744\pm0.129$ & 0.017  \\ 
0.05 & $1.472\pm0.103$ & 0.027  &$0.463\pm0.043$ & 0.0072 &$0.327\pm0.043$ & 0.022  \\ 
0.10 & $0.666\pm0.049$ & 0.026  &$0.340\pm0.036$ & 0.0077 &$0.214\pm0.025$ & 0.022  \\ 
0.15 & $0.528\pm0.037$ & 0.027  &$0.300\pm0.028$ & 0.0081 &$0.180\pm0.020$ & 0.021  \\ 
0.20 & $0.524\pm0.040$ & 0.027  &$0.279\pm0.027$ & 0.0086 &$0.176\pm0.019$ & 0.022  \\ 
0.25 & $0.536\pm0.040$ & 0.027  &$0.315\pm0.034$ & 0.0086 &$0.195\pm0.020$ & 0.022  \\ 
0.30 & $0.546\pm0.039$ & 0.027  &$0.297\pm0.027$ & 0.0088 &$0.206\pm0.020$ & 0.023  \\ 
0.35 & $0.534\pm0.039$ & 0.027  &$0.322\pm0.030$ & 0.0095 &$0.189\pm0.020$ & 0.024  \\ 
0.40 & $0.571\pm0.036$ & 0.028  &$0.319\pm0.027$ & 0.0099 &$0.212\pm0.023$ & 0.023  \\ 
0.45 & $0.549\pm0.038$ & 0.029  &$0.351\pm0.028$ & 0.0102 &$0.213\pm0.020$ & 0.025  \\ 
0.50 & $0.598\pm0.038$ & 0.030  &$0.372\pm0.031$ & 0.0099 &$0.249\pm0.023$ & 0.025  \\ 
0.55 & $0.623\pm0.043$ & 0.030  &$0.448\pm0.041$ & 0.0101 &$0.267\pm0.024$ & 0.025  \\ 
0.60 & $0.706\pm0.048$ & 0.031  &$0.494\pm0.038$ & 0.0104 &$0.277\pm0.024$ & 0.025  \\ 
0.65 & $0.776\pm0.051$ & 0.031  &$0.584\pm0.043$ & 0.0107 &$0.354\pm0.029$ & 0.027  \\ 
0.70 & $0.856\pm0.056$ & 0.031  &$0.688\pm0.044$ & 0.0114 &$0.381\pm0.028$ & 0.026  \\ 
0.75 & $0.973\pm0.060$ & 0.032  &$0.831\pm0.059$ & 0.0118 &$0.566\pm0.038$ & 0.028  \\ 
0.80 & $1.119\pm0.071$ & 0.032  &$1.067\pm0.069$ & 0.0117 &$0.774\pm0.050$ & 0.027  \\ 
0.85 & $1.235\pm0.081$ & 0.032  &$1.570\pm0.094$ & 0.0121 &$1.231\pm0.063$ & 0.029  \\ 
0.90 & $1.302\pm0.095$ & 0.032  &$2.180\pm0.135$ & 0.0127 &$2.288\pm0.097$ & 0.030  \\ 
0.95 & $1.200\pm0.107$ & 0.032  &$3.355\pm0.169$ & 0.0130 &$4.480\pm0.143$ & 0.029  \\ 
1.00 & $1.068\pm0.128$ & 0.032  &$3.392\pm0.248$ & 0.0138 &$6.477\pm0.256$ & 0.031  \\
\enddata
\tablecomments{The flux PDF, $P(F)$, and the average rms noise per 
pixel, $n(F)$, are averaged over bins covering
the flux range within $\Delta F=\pm0.025$ of 
the listed value of $F$.  The first and last bins 
include the few additional points with $F<-0.025$ and 
$F>1.025$, respectively.}
\end{deluxetable}  
We present here the probability distribution function (hereafter, PDF)
of the transmitted flux, first used as a tool to study the \lya forest
by \citet{jo91}. The PDF of the same observations used
here (except for our addition of an eighth quasar in the sample) was
presented before in \citet{rms97}. The results will be given here in
differential form and with error bars computed as described in Appendix
A. The PDF was measured from the observations with all pixels weighted
equally. Table \ref{pdftab} gives $P(F)$ and the noise amplitude $n(F)$,
which is necessary for comparisons with theory.
We define $n(F)$ to be the average of $\sigma_i$ over all the 
pixels in each flux bin, where $\sigma_i$ is the noise in each pixel.
The noise increases with increasing $F$ (note that the $z=3$ data has 
significantly less noise than the other two redshift bins).
We use 21 bins of width $\Delta F=0.05$ with the first centered on 
$F=0$ and the last on $F=1$.  Pixels with flux greater (less) than
$F=1.025$ ($F=-0.025$) are included in the last (first) bin. 
We provide statistical error bars on the
probabilities to facilitate comparisons with theory but it is 
important
to recognize that these error bars are significantly correlated.  
If a $\chi^2$ statistic is computed from only these diagonal elements 
of the
error matrix the distribution will be approximately twice as wide
as the distribution of $\chi^2$ properly computed from the full error
matrix. The error matrix is available in the website quoted in the
introduction.

\begin{figure}
\plotone{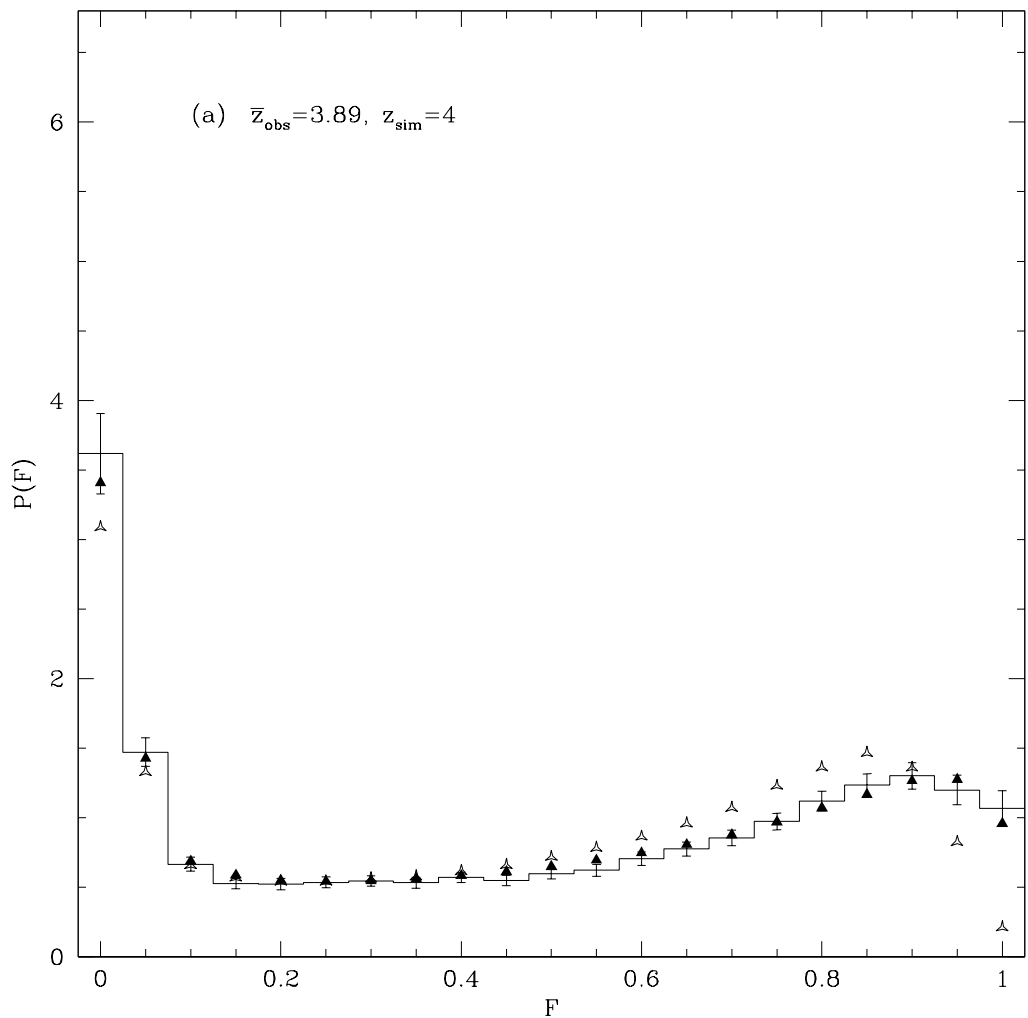}
\caption{The PDF of $F$ for the observations
({\it histograms}) and for
the simulation with the continuum fitting approximation 
({\it filled points}) and without it ({\it open points}).
The small number of points outside the displayed range of $F$ are 
included in the outermost bins.
Errors bars were generated by bootstrap resampling.   
The numerical simulation has $\bar{F}$ fixed to agree with
the observations.   (a) shows $\bar{z}=3.89$, (b) shows 
$\bar{z}=3.00$, and (c) shows $\bar{z}=2.41$.
}
\label{pdf3}
\end{figure}  
\begin{figure}
\plotone{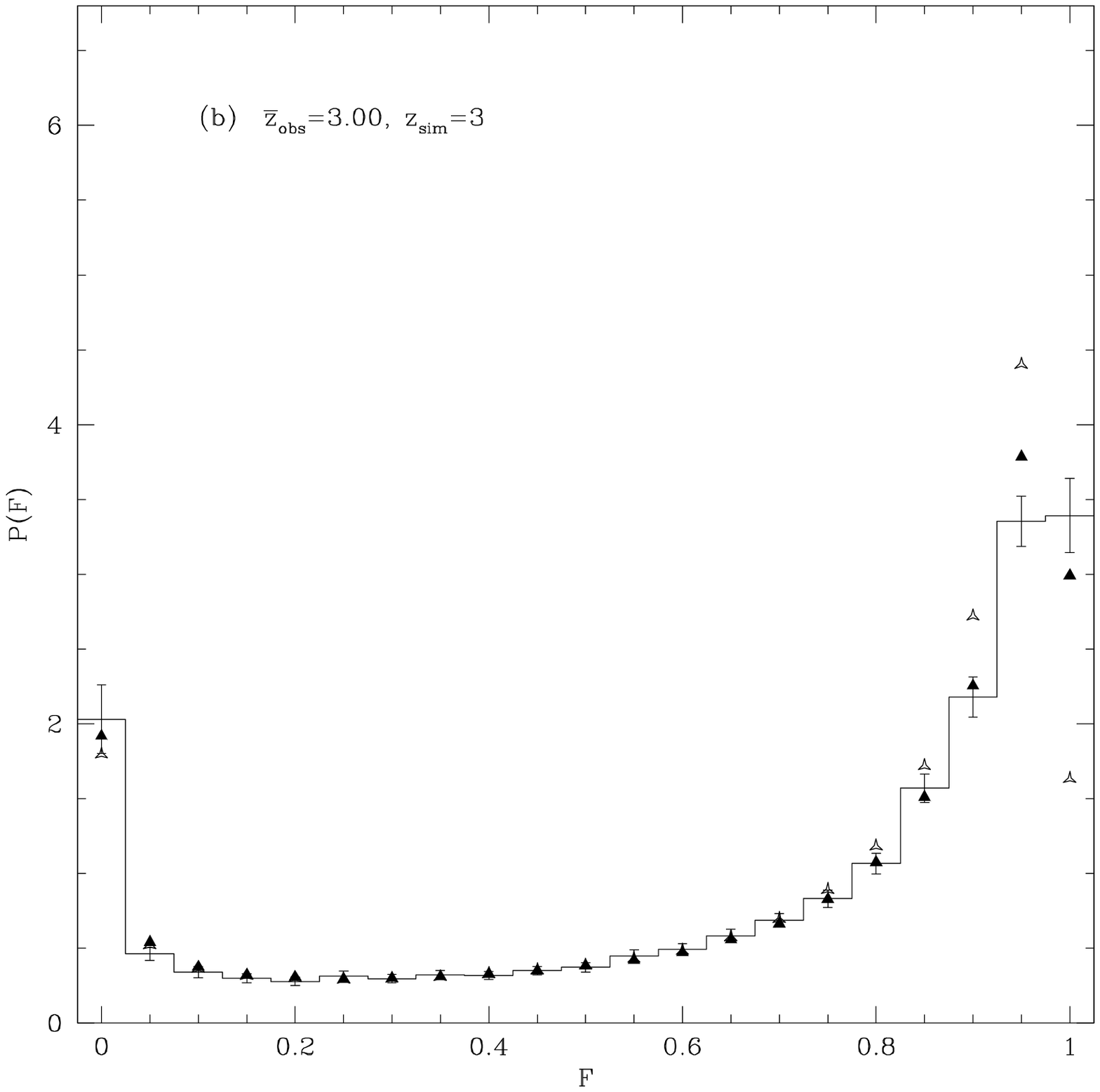}
\end{figure}  
\begin{figure}
\plotone{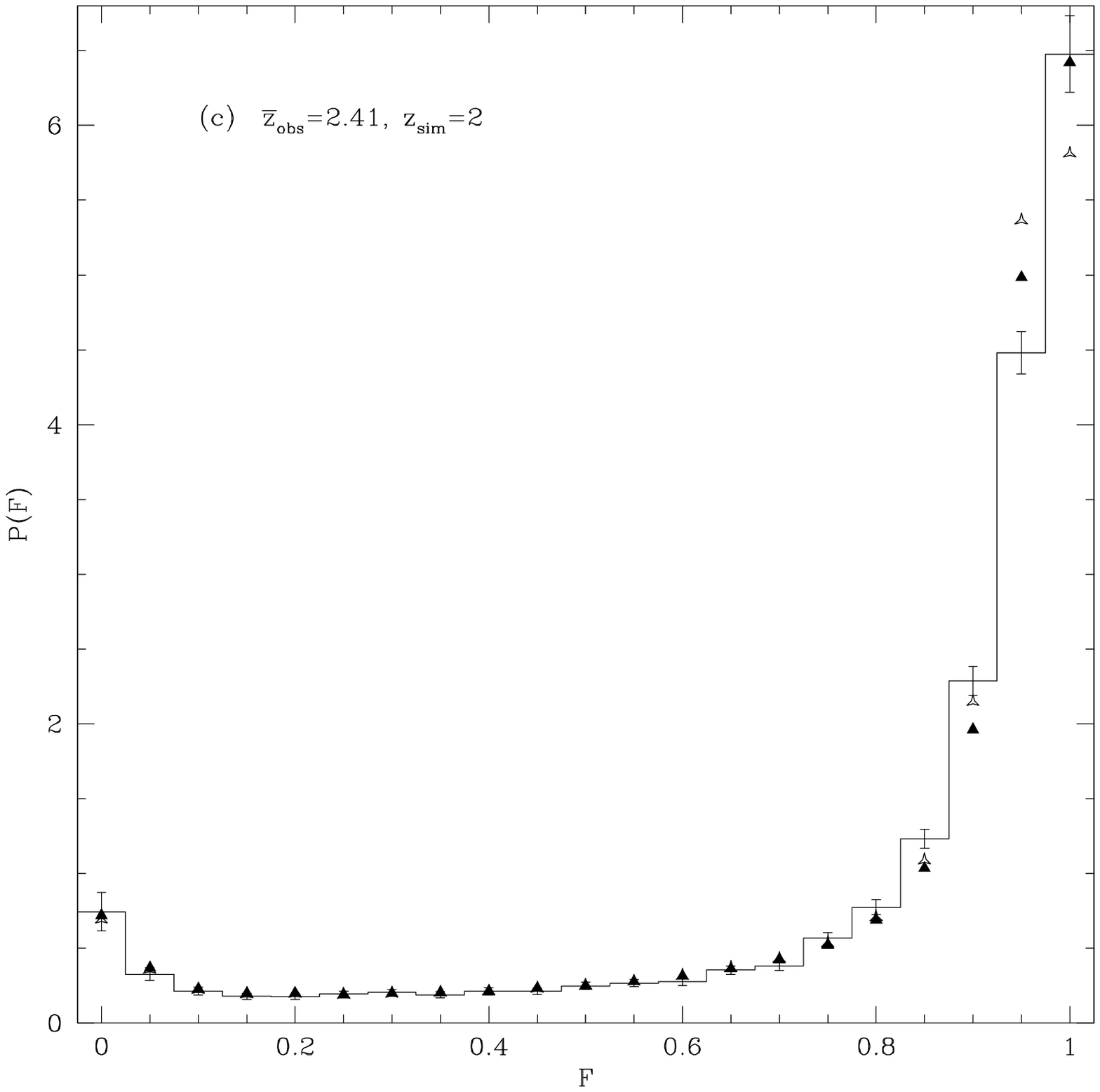}
\end{figure}  
Figures \ref{pdf3}(a,b,c) show the PDF of the 
observations and the simulation.  The mean transmitted flux in the
simulation is matched to the observed mean.  We compute a $\chi^2$
statistic (using the full error matrix) to compare the observations 
and simulation. The results are
$\chi^2/\nu=1.1$ (with $\nu=19$, 39\% likelihood) for $z= 3.89$,
$\chi^2/\nu=1.7$ (3\% likelihood) for $z= 3$, and
$\chi^2/\nu=4.4$ (negligible likelihood) for $z= 2.41$.  
Figures \ref{pdf3}(a,b,c) show the effect of our continuum fitting
approximation on the simulated PDF.  This correction is important to
the agreement seen in the two higher redshift comparisons.

  The almost perfect agreement of the predicted flux distribution in the
simulation and the observed distribution is impressive, and is one of
the
strong pieces of evidence in favor of the new \lya forest theory based
on gravitational evolution of primordial fluctuations.
We believe that even the small disagreement between the simulation and 
observations at $\bar{z}=2.41$ (Figure \ref{pdf3}(a)) is due primarily 
to the imperfection of our continuum fitting approximation.
Since the inclusion of the continuum fitting correction moves the
predicted PDF in the direction of the observed one, as seen in Figure
\ref{pdf3}, it is plausible that the small remaining discrepancy is
due to an underestimate of this correction. We have examined other
possible reasons for this discrepancy, and they were all found to be
not significant. One of these is that the noise that we
include in the simulations (assumed to have a Gaussian distribution for
a given flux $F$, obtained from the average of the variance of all
pixels with flux $F$ in Table \ref{pdftab}) does not adequately
represent the true distribution function of the noise, which is
actually a sum of Gaussians with the variance distribution of all the
pixels with flux $F$ in the observations. We tested this by
computing the predicted $P(F)$ given the actual distribution of noise,
and found that the difference it makes is much smaller than the
differences between prediction and observation at $\bar{z}=2.41$. Thus,
the average noise values we give in Table \ref{pdftab} should be
sufficient to compare results of any other numerical simulations with
the observational data presented here on $P(F)$. 
Another effect is that observational results are obtained by averaging
over a certain redshift interval, whereas the simulation predictions
are for a fixed redshift. The importance of this effect was tested by
creating sets of spectra where each line of sight had a different value
for $\mu$, given by $\mu=\mu_0~[(1+z)/(1+\bar{z})]^{4.5}$, where $\mu_0$ 
is the value of $\mu$ at the mean redshift and $z$ is varied across 
the full width of the redshift bin in the observational data; there was
negligible change in the predicted PDF of the flux.

  Why is there such a good agreement between the predicted $P(F)$ and
the observed one, after having adjusted only one parameter (the mean
transmitted flux)? To understand the significance of this result, it is
useful to think of the optical depth on a given pixel as being
determined mostly by the gas density and temperature at a given point
in space, $\tau \propto \rho^2 T^{-0.7}$ (from photoionization
equilibrium). Given a $\rho-T$ relation determined by the
photoionization history \citep{hg97}, the predicted $P(F)$
should essentially be a result of gravitational evolution starting
from Gaussian initial conditions, and it should mostly depend on one
parameter only (in addition to the mean transmitted flux):
the amplitude of fluctuations on the Jeans scale.
Obviously, the higher the amplitude of density fluctuations, the larger
the dispersion in the transmitted flux should be. A weaker dependence
on the $\rho-T$ relation can also be expected. This suggests two
possible
implications of the good agreement of the prediction of $P(F)$ with the
observations, which will need to be examined further: (a) the \lya 
forest is indeed a result of gravitational
evolution in a photoionized IGM starting from Gaussian initial
conditions; (b) the amplitude of fluctuations in the $\Lambda$CDM model
assumed in the simulation is close to the true value in the universe.

  In the next Section, our analysis of the power spectrum will show
that the amplitude of fluctuations in the simulation we use should
actually be reduced by $\sim 15\%$ to match the observations. However,
the finite size of the simulated box reduces the effective power,
resulting by chance in a value of the variance of the transmitted flux
that matches very well the observations (see \S\ref{corfuncsection}).

\section{THE POWER SPECTRUM}\label{powspecsection}

In this section we compute the one dimensional power spectrum of the 
transmitted flux, $P_F(k)$. The flux power spectrum is the most 
straightforward two-point statistic that can be measured from the data.
Hopefully this will make the results more generally useful for
comparisons with analytic theory, numerical simulations, and other
observations. We shall then study the relation of the flux power
spectrum to the linear mass power spectrum using the numerical
simulation, on scales that are large enough to make the fluctuations in
\lya absorption be related to linear density fluctuations. This relation
in the large-scale limit is further discussed in Appendix C.
Although there are theoretical reasons to Gaussianize the
transmitted flux before obtaining the power spectrum \citep{cwk98}
this operation can amplify the effect of the noise, and its merit in
improving the recovery of the linear power spectrum has not been made
clear.

The data is given in the form of pixels with wavelength label 
$\lambda_i$ and flux value $F_i$. We measure distance between pixels in
units of the local velocity scale using the formula
\begin{equation}
\Delta v_i= \frac{H(\bar{z})}{1+\bar{z}}\Delta r_i =
\frac{H(\bar{z})}{1+\bar{z}} 
\int_{\bar{z}}^{z_i}\frac{c~dz'}{H(z')}= 
2 c \left(1-\sqrt{\frac{\bar{\lambda}}{\lambda_i}}
\right)~,
\end{equation}
where $\bar{\lambda}=\lambda_\alpha (1+\bar{z})$ is the wavelength at 
the mean redshift, $\bar{z}$, of any given data subset, and
$\Delta r_i$ is the comoving distance between pixel $i$ and
a pixel at the mean redshift, where we have assumed an Einstein-de 
Sitter universe for this calculation.
With this formula, separations are precisely proportional to comoving
distance for an Einstein-de Sitter universe, and are a close enough
approximation for other cosmologies for the redshifts intervals we 
shall use. The power spectrum is then estimated
from each spectrum using the Lomb periodogram code in
\citet{ptv92}.  This algorithm is designed for use on unevenly
sampled data (the uneven sampling in our case is due to the removal of
the chunks in the spectra containing damped \lya and metal lines,
and to the change in pixel size with redshift),
and avoids rebinning of the data.
The computed modes are averaged over bins evenly spaced in
$\log(k)$.

\subsection{Results for the Observed $P_F(k)$}

\begin{deluxetable}{lllll}
\tablecolumns{5}
\tablecaption{The observed power spectrum of the flux.
\label{pspectab}}
\tablehead{
\colhead{$k_{bin, min}$} & \colhead{$k_{mean}$} & 
\colhead{$P_F(k)$ $(\bar{z}=3.89)$} & 
\colhead{$P_F(k)$ $(\bar{z}=3.00)$} & \colhead{$P_F(k)$
$(\bar{z}=2.41)$} \\
($[\kms]^{-1}$) & ($[\kms]^{-1}$) & ($\kms$) & ($\kms$) & ($\kms$) }
\startdata
0.00251 & 0.00284 & $17.1    \pm2.8    $ & $20.8    \pm3.9    $ & $9.17    \pm2.6    $ \\ 
0.00316 & 0.00358 & $12      \pm4      $ & $16.1    \pm4      $ & $9.68    \pm2.6    $ \\ 
0.00398 & 0.0045  & $18.9    \pm3.3    $ & $14.7    \pm2.8    $ & $13.3    \pm3.4    $ \\ 
0.00501 & 0.00566 & $17.5    \pm4.7    $ & $22.7    \pm4.7    $ & $11.4    \pm2.2    $ \\ 
0.00631 & 0.00713 & $16.9    \pm3.1    $ & $10.8    \pm1.6    $ & $11.4    \pm1.9    $ \\ 
0.00795 & 0.00898 & $15.4    \pm2.4    $ & $10.1    \pm1.7    $ & $10.2    \pm0.95   $ \\ 
0.01    & 0.0113  & $9.44    \pm1.1    $ & $9.08    \pm1.9    $ & $7.5     \pm1.6    $ \\ 
0.0126  & 0.0142  & $8.26    \pm1.4    $ & $8.07    \pm1      $ & $5.95    \pm0.9    $ \\ 
0.0159  & 0.0179  & $8.15    \pm1.1    $ & $5.61    \pm0.35   $ & $4.11    \pm0.5    $ \\ 
0.02    & 0.0225  & $4.88    \pm0.57   $ & $4.96    \pm0.7    $ & $3.58    \pm0.35   $ \\ 
0.0251  & 0.0284  & $4.12    \pm0.34   $ & $3.59    \pm0.43   $ & $2.5     \pm0.31   $ \\ 
0.0316  & 0.0357  & $3.03    \pm0.3    $ & $2.11    \pm0.3    $ & $1.49    \pm0.16   $ \\ 
0.0398  & 0.045   & $1.94    \pm0.24   $ & $1.53    \pm0.14   $ & $1.05    \pm0.077  $ \\ 
0.0501  & 0.0566  & $1.12    \pm0.15   $ & $0.896   \pm0.094  $ & $0.555   \pm0.038  $ \\ 
0.0631  & 0.0713  & $0.629   \pm0.049  $ & $0.435   \pm0.051  $ & $0.28    \pm0.02   $ \\ 
0.0795  & 0.0898  & $0.349   \pm0.033  $ & $0.188   \pm0.016  $ & $0.136   \pm0.014  $ \\ 
0.1     & 0.113   & $0.141   \pm0.013  $ & $0.0691  \pm0.007  $ & $0.0473  \pm0.0044 $ \\ 
0.126   & 0.142   & $0.0499  \pm0.0041 $ & $0.0208  \pm0.0022 $ & $0.0217  \pm0.0025 $ \\ 
\enddata
\tablecomments{Averaged over bins
defined by the listed values of $k_{min}$
[the maximum $k$ for the final bin is $k_{max}=0.159$ $(\kms)^{-1}$],
with average $k$ value $k_{mean}$. 
}
\end{deluxetable}  
Table \ref{pspectab} lists the results of the flux power spectrum 
measurements from the observational data, over the range $0.0025 ~
(\kms)^{-1} < k < 0.16 (\kms)^{-1}$ (the power at
$k < 0.0025 (\kms)^{-1}$ is possibly distorted by the
continuum fitting operation, and at $k > 0.16 (\kms)^{-1}$ the power
is strongly affected by narrow metal lines and other systematic
errors that are discussed below).
Note that the velocity scales represent different comoving
scales at each of the three redshifts.
The error bars were determined as described in Appendix B; for the
power spectrum they are approximately independent (as expected in
linear theory). Our normalization convention is that the rms flux
fluctuation is $\sigma_F^2 = \int_{-\infty}^{\infty} (dk/2\pi)\,
P_F(k)$.

\begin{figure}
\plotone{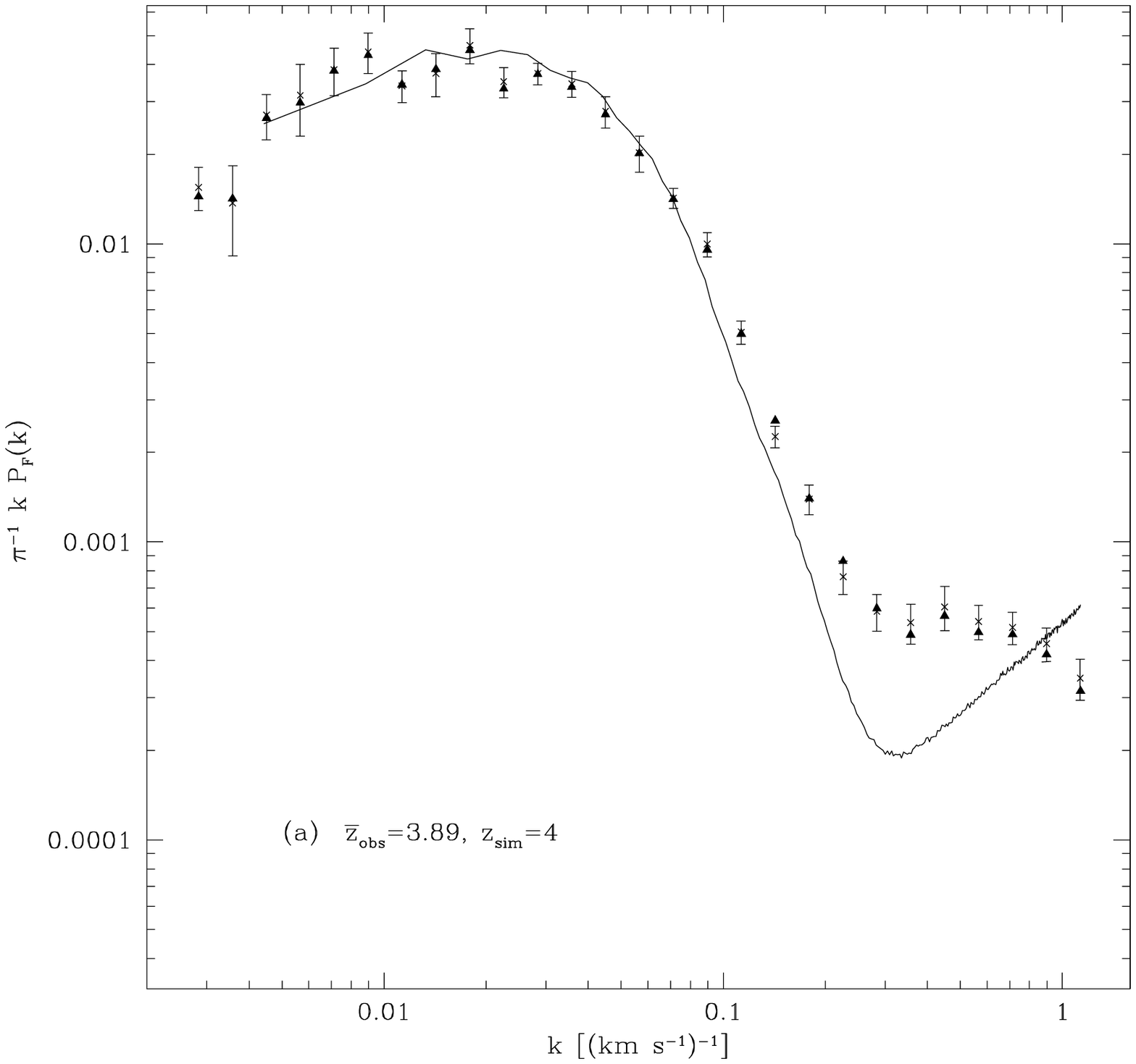}

\caption{The observed one dimensional power spectra computed
with and without including the regions possibly contaminated by 
metal lines.  The points with error bars show $P_F(k)$ 
computed
after excluding the possibly contaminated regions while the triangles
show the points computed from the complete spectra.  The solid 
lines show the power spectra from the simulation with the
mean transmitted flux adjusted to match the observations.
(a) shows $\bar{z}=3.89$, (b) shows 
$\bar{z}=3.00$, and (c) shows $\bar{z}=2.41$.}
\label{ps3}
\end{figure}
\begin{figure}
\plotone{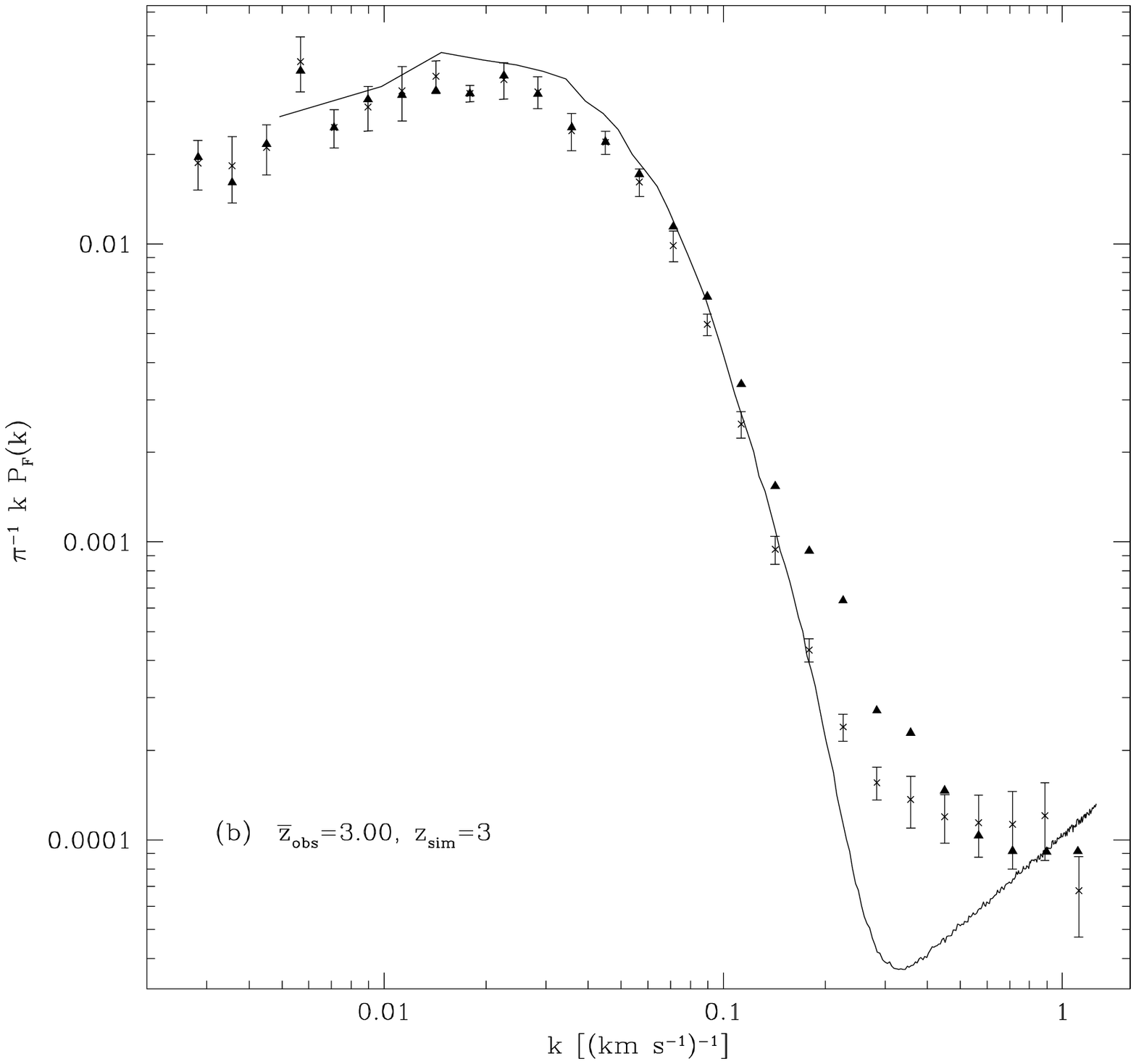}
\end{figure}
\begin{figure}
\plotone{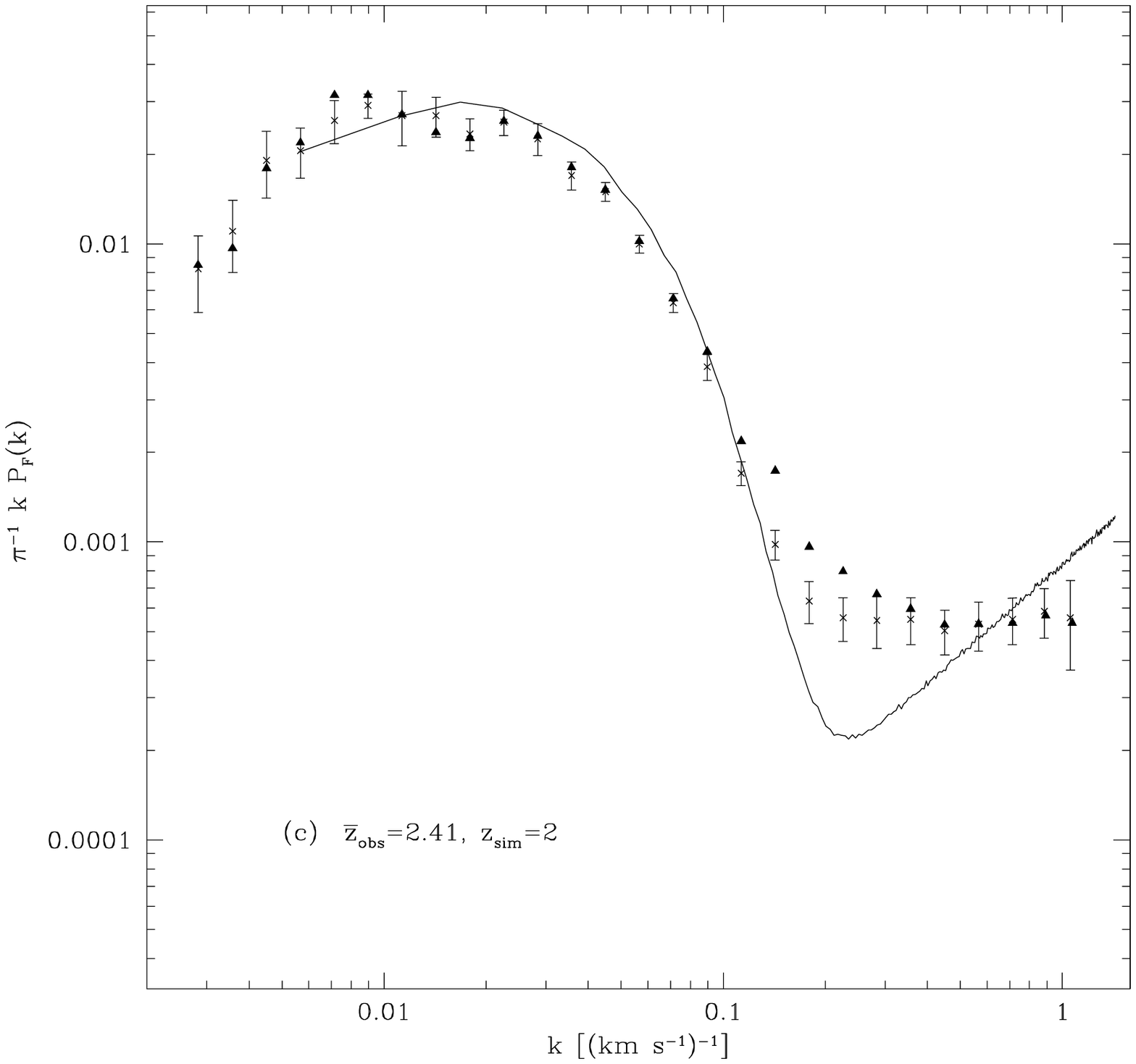}
\end{figure}
  The observed $P_F(k)$ is plotted in Figures \ref{ps3}(a,b,c)
(crosses with error bars) for the three usual redshift bins. Notice 
that
we plot the quantity $k\, P_F(k)$, for easier visualization. We also
plot the power spectrum measured from the simulation (solid line), 
with
the mean flux decrement adjusted to match the observation in every
redshift interval.
There is generally very good qualitative agreement between the
simulations and observations on large scales
($k \lesssim 0.1 (\kms)^{-1}$).
On small scales ($k\gtrsim 0.3 (\kms)^{-1}$), the simulated power
spectrum is constant due to the noise that we add, matching that in the
observations. The different behavior of the observed $P_F(k)$ is due
to various effects that we shall now describe.

  The measurement of the power spectrum is complicated by the possible
presence of metal lines in the spectra. Because the metal lines are
narrower than the \lya forest lines, they can affect the power
spectrum on very small scales. The numbers listed
in Table \ref{pspectab}, shown as crosses in Figure \ref{ps3}, exclude
potentially contaminated regions, but some of these regions might
contain genuine \lya lines that were selectively eliminated as 
suspected
metal lines because they are narrow.
The solid triangles in Figure \ref{ps3} show the power spectrum with
these regions included (except for a few regions that contain damped 
\lya systems or bad data points).
We see that the suspected metal lines are significant in adding small
scale power. This problem in the determination
of the power spectrum occurs on the smallest scales only (starting
with the fifteenth bin in Table \ref{pspectab}).

The removal of chunks of spectra suspected of containing 
metal lines changes the effective window function for the power 
spectrum measurement.
\begin{figure}
\plotone{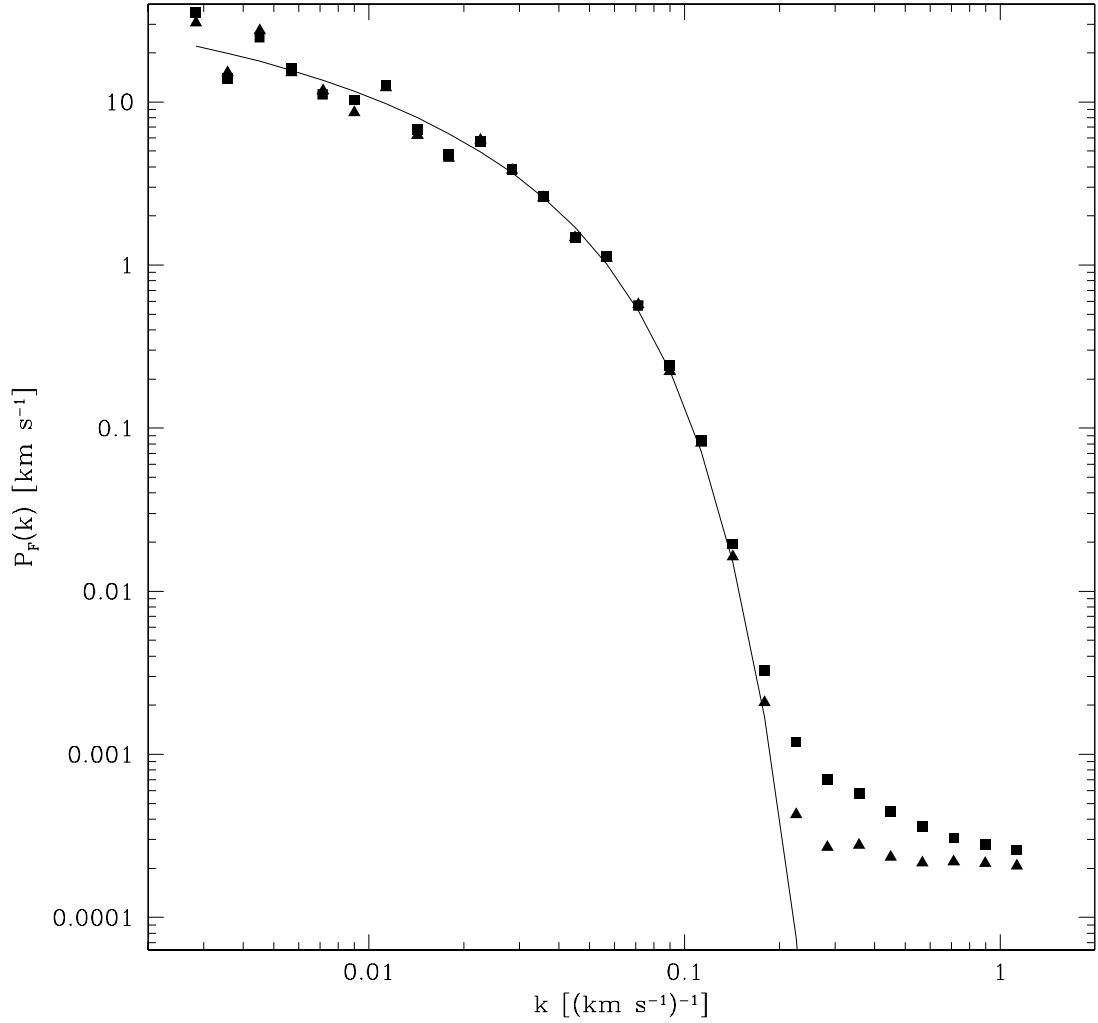}
\caption{ The power spectrum measured from
randomly generated spectra with input power spectrum shown by the solid
curve (see Appendix A).  The triangles show the power measured from 
the full spectrum,
while the squares show the power measured after the regions that are
suspect in the real data are removed.  The power is 
increased on the smallest scales when the window function is modified
by removing the chunks.  Figure \ref{ps3} shows the opposite effect
because the removed regions in the real data contain narrow lines.
The flattening of both sets of points on the smallest scales is because
of the noise added to the random spectra.}
\label{compmycutsim}
\end{figure}
To show the magnitude of this effect, we have randomly generated 
Gaussian spectra with the one-dimensional, linear power spectrum shown 
by the solid line in Figure \ref{compmycutsim}, with a total length in 
velocity equal to the size of the observational sample used here in 
each redshift interval. The procedure we have used for
generating these mock spectra is described in detail in Appendix A.
We have then measured the power spectrum with the same method used for
the observations, with and without removing the same chunks of the
spectra that contain metal lines in the real observations.  
The measured $P_F(k)$ points for the case where the chunks are
not removed (solid triangles) appear to follow the input power spectrum
(except for statistical fluctuations at small $k$), until the power is
small enough that the added noise begins to dominate. When the chunks
are removed (solid squares), the change in the window function results
in increased power on small scales. In the observational data the
removed chunks contain narrow lines which have a greater effect on the
power spectrum than the change in the window function, so the power
decreases when the chunks are removed, except for $z_{obs}=3.89$ where
the two effects approximately cancel.

  In the rest of \S\ref{powspecsection}, we do a quantitative comparison of the observed
and simulated power spectrum. We will not use the measurements on small
scales where the effects of the metal lines and the window function
are important.

\subsection{Fitting Formula for $P_F(k)$}

The mass power spectrum of the linear density perturbations for cold
dark matter models can be fitted by the form
\begin{equation}
P_{3D}(q) = A\, \frac{q^n [\ln(1+\alpha_1 q)/\alpha_1 q]^2}
         {[1+\alpha_2 q +(\alpha_3 q)^2+(\alpha_4 q)^3 +
	  (\alpha_5 q)^4]^{\frac{1}{2}}}~,
	  \label{pspecformula3D}
\end{equation}
where $q\equiv k/\Gamma(z)$, $\Gamma(z) = (1+z) \Omega_0 h^2 /H(z)$
(when $k$ has units of inverse velocity), and $A$ is a normalization
constant. The subscript $3D$ is written
here to remind us that this is the power spectrum for fluctuations in
three-dimensional real space.
The formula is given by \citet{bbk86}, but we modify the 
parameters to the fit for $\Omega_b=0.05$: 
$\alpha_1=2.205$, $\alpha_2=4.05$, $\alpha_3=18.3$, $\alpha_4=8.725$,
and $\alpha_5=8.0$ \citep{m96}.
The amplitude parameter $A$ has units of $(\kms)^3$.
For the initial conditions of our simulation, $n=0.95$, $\Gamma(z)$
can be found using $\Omega_0=0.4$, $\Omega_\Lambda=0.6$, and $h=0.65$,
and then $A$ can be found from $\sigma_8=0.79$.

  \citet{cwk98} found that multiplying the three dimensional
linear theory power spectrum by the smoothing function 
$\exp(-k^2 w_c^2)$, with
$w_c \simeq 34\kms$ at $z=3$, matches the 
\lya forest power spectrum obtained in their simulations.
We also find that this function fits the output of our simulation 
reasonably well down to remarkably small scales, although we obtain 
smaller values 
of the 3D Gaussian cutoff: $w_c\simeq 12\kms$.
It seems probable that, as \citet{cwp99} speculated, the high 
$w_c$ found by \citet{cwk98} is a result of their lower simulation 
resolution.  Their
mean particle separation was close to $34 \kms$.

  Here, we will fit the one-dimensional flux power spectrum in the
observations and simulation with the following formula: 
\begin{equation}
    P_F(k)=A_F\, \frac{\exp(-k^2 v_c^2)}{2\pi}
          \int_{k}^\infty dk' \, k' P_{3D}(k') ~,
	  \label{pspecformula}
\end{equation}
where $P_{3D}(k)$ is given by equation (\ref{pspecformula3D}). 
We will
first obtain fits with the two free parameters $A_F$ and $v_c$, and then
add the slope $n$ in \S\ref{slopenovaramp}. 
Equation (\ref{pspecformula}) applies a
Gaussian smoothing directly to the 1D power spectrum, instead of
smoothing in three dimensions and then converting to the one-dimensional
spectrum \citep[as done in][]{cwp99}. We will find later 
(in Fig. \ref{slopechisqsim})
that equation (\ref{pspecformula}) agrees well with the low-$k$
[$<0.04~(\kms)^{-1}$] flux power spectrum of the simulation and, most 
importantly, the best fit for the slope parameter coincides with
the real space linear theory slope predicted from the initial 
perturbations.
While the 3D Gaussian smoothing provides a better fit for the power
spectrum on very small scales than the 1D Gaussian smoothing,
we have found that the best fit parameters with the 3D smoothing
do not return the correct value of the spectral index $n$ of the model
used in the simulation (the basic reason is that the effect of the 3D
smoothing on $P_F(k)$ extends to larger scales than the 1D smoothing).
This is not inconsistent with \citet{cwk98}, because
we extend the fit to smaller scales.

  In Appendix C, we give a justification of the fitting formula in 
eq.\ (\ref{pspecformula}) in the limit of large scales, showing that
the flux power spectrum must be proportional to the one-dimensional
mass power spectrum \citep[see also][]{sw98}.

  Notice that in strict linear theory we should include the effects of 
peculiar velocities in the transformation from three to one dimensional
power, which affect the shape of the power spectrum
\citep{h99,mm99a}. However, these linear
theory results work only on extremely large scales, and in the range of
scales we explore here (where the density fluctuations are not much
smaller than unity), the higher-order effects are important and opposite
to the linear effects. Like \citet{cwk98}, we find that the slope
of the simulation power spectrum on large scales is consistent with the
{\it real space} linear theory slope (or equivalently, the peculiar
velocity parameter $\beta$ is quite small).

  We first fit equation (\ref{pspecformula}) to the observations and
simulations leaving as free parameters the amplitude $A$ and the
Gaussian cutoff $v_c$ only. All the other parameters are held fixed
at the values they have in the simulation in linear theory (we will
let the spectral index $n$ vary later, in \S 5.5).
\begin{deluxetable}{ccccccc}
\tablecolumns{7}
\tablecaption{Fitted parameter values for equation 
(\ref{pspecformula}).
\label{fittab}}
\tablehead{
\colhead{$P_F(k)$ from:} & \colhead{$z$} & \colhead{$\bar{F}$} & 
\colhead{$\Delta_F^2(k_p)$} & \colhead{$v_c$} & 
\colhead{$\chi^2/\nu$} & \colhead{$\nu$} \\ 
&&&&($\kms$)&&}
\startdata
obs. & 3.89 & 0.475 & $0.0392\pm 0.0024$ & 21.6 & 1.2 & 10 \\
obs. & 3.00 & 0.684 & $0.0370\pm 0.0021$ & 25.4 & 0.77 & 10 \\
obs. & 2.41 & 0.818 & $0.0321\pm 0.0021$ & 28.8 & 1.4 & 9 \\
sim. & 4 & 0.475 & $0.0436\pm 0.0010$ & 21.9 & 2.2 & 5 \\
sim. & 4 & 0.684 & $0.0370\pm 0.0009$ & 20.6 & 1.9 & 5 \\
sim. & 4 & 0.818 & $0.0247\pm 0.0006$ & 19.2 & 2.6 & 5 \\
sim. & 3 & 0.475 & $0.0511\pm 0.0014$ & 23.8 & 1.4 & 5 \\
sim. & 3 & 0.684 & $0.0442\pm 0.0012$ & 23.0 & 1.3 & 5 \\
sim. & 3 & 0.818 & $0.0290\pm 0.0007$ & 21.5 & 1.6 & 5 \\
sim. & 2 & 0.475 & $0.0635\pm 0.0019$ & 29.4 & 0.36 & 3 \\
sim. & 2 & 0.684 & $0.0544\pm 0.0018$ & 28.7 & 0.26 & 3 \\
sim. & 2 & 0.818 & $0.0347\pm 0.0012$ & 26.7 & 0.11 & 3 \\
\enddata
\tablecomments{  
The fitted power spectrum amplitude is given in
terms of $\Delta_F^2(k_p)=(2 \pi^2)^{-1} k_p^3 P_{F3D}(k_p)$ where
$k_p=0.04~(\kms)^{-1}$ at $z=3$ (and is held constant in comoving 
coordinates).
}
\end{deluxetable} 
Table \ref{fittab} gives the best fit values of these two parameters.
For each simulation output (at redshift $z=2,3,4$) we perform three
separate fits using the mean flux decrements $\bar{F}=0.475$, 0.684,
and 0.818. 
The result for the amplitude parameter is listed in terms of the 
contribution to the variance per unit interval of $\ln{k}$,
$\Delta_{3D}^2(k_p)=(2 \pi^2)^{-1} k_p^3 P_{3D}(k_p)$ [where 
$P_{3D}(k)$ is given by eq.\ (\ref{pspecformula3D}) with the
value of $A$ to fit $P_F$ in eq.\ (\ref{pspecformula}) ].
The value of $k_p$ is $k_p=0.04~(\kms)^{-1}$ at $z=3$, and is held 
constant in comoving coordinates.
The reason for this choice of $k_p$ will be made clear below, in
\S 5.6.

  In order to perform the same fits to the flux power spectrum predicted
by the simulation, we have generated approximate error bars by dividing
the lines of sight through the simulation box into twelve groups and
measuring the dispersion among them. The groups are defined by the four 
quadrants of each of the three faces of the simulation cube.
We expect that this is an underestimate of the actual errors that 
would be found if many simulation boxes were available because the
twelve groups are not independent, but these error bars should give
reasonable best values for fitted parameters.

  Even though the power spectrum of the optical depth should have a
sharp cutoff at the thermal broadening scale, the transformation
to transmitted flux creates small scale power, so the Gaussian cutoff
cannot be expected to fit on small scales. We choose the maximum $k$
values to use in the fits $(k_{max})$ by estimating where equation
(\ref{pspecformula}) begins to fit the simulation outputs poorly.
We use $k_{max}=0.04~(\kms)^{-1}$ for $z_{obs} = 3$ and
$z_{obs}= 3.89$, and $k_{max}=0.032~(\kms)^{-1}$ for $z_{obs} = 2.41$.
The cutoff $k_{max}$ corresponds roughly to
$v_c^{-1}$ and decreases with time because $v_c$ increases with time.

The values of $\chi^2$ for the fits all have probabilities higher
than $5\%$ with the exception of the fit to the $z=4$ output of the
simulation with $\bar{F}=0.818$.  The probability of $\chi^2$ exceeding
its value for the worst fit 
is $\sim$2\%, but this is 
probably acceptable because our simulation
error bars are likely to be slightly underestimated.

\subsection{Redshift Evolution of the Power Spectrum}

  If the \lya forest is in fact governed by gravitational evolution,
we know that the amplitude of the fluctuations should generally
increase with time. In the linear regime, and when $\Omega(z) \simeq
1$ (a very good approximation in the model we use at $z>2$),
the amplitude should grow proportionally to the scale factor. However,
when analyzing the flux power spectrum, its amplitude is affected by
the mean flux decrement which is also evolving with redshift.

\begin{figure}
\plotone{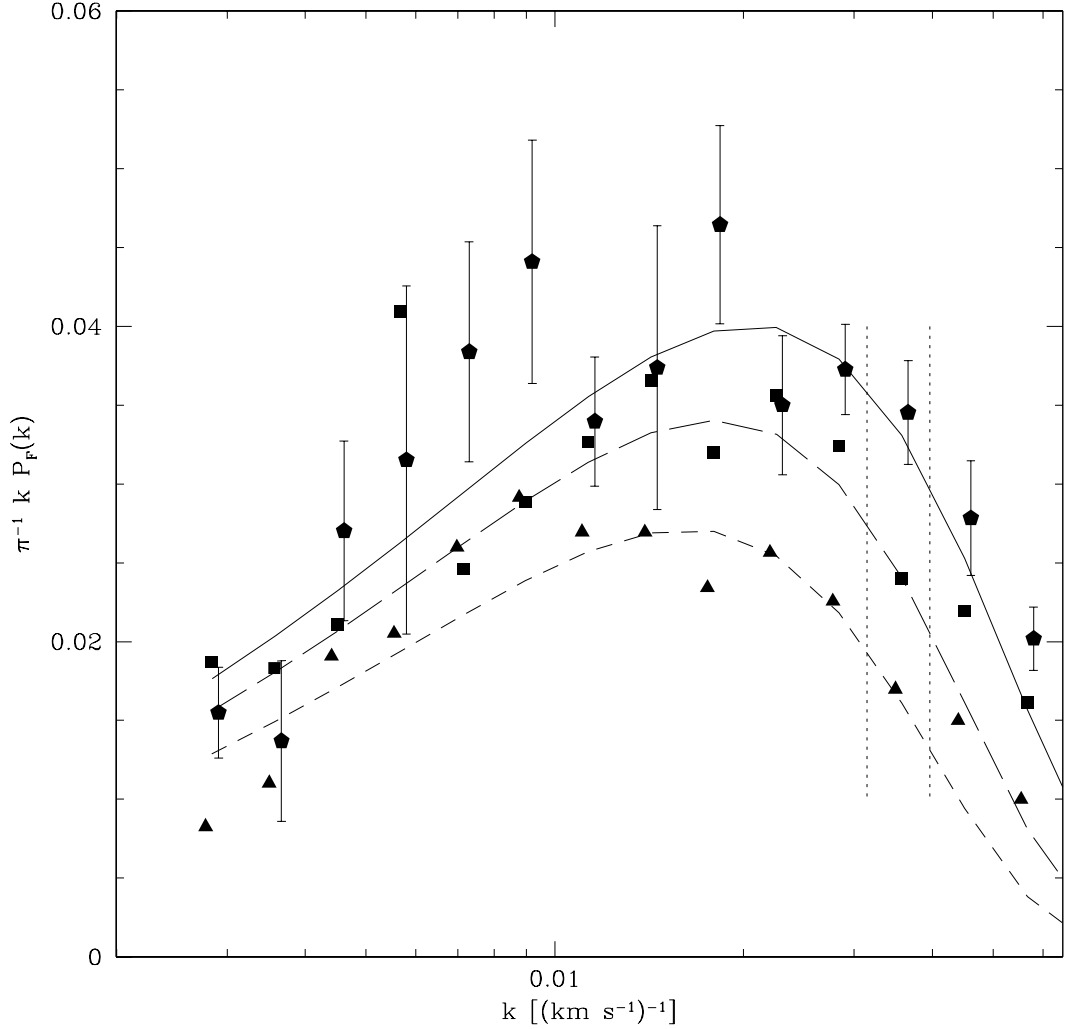}
\caption{The observed one dimensional power spectrum of $F$ along the 
line of sight for three redshift intervals.  The points show the 
observed $P_F(k)$ values while the lines are analytic fits to the 
points.
The redshifts $\bar{z}=3.89$, $3.0$, and $2.41$ are symbolized by 
pentagons and the solid line, squares and the dotted line, and 
triangles and the dashed line, respectively.  The power is reduced 
with
decreasing redshift because of the change in the mean flux decrement.
The fits are obtained using only points at $k$ smaller than the
left vertical dotted line for $\bar z=2.41$, and the right vertical
dotted line for $\bar z = 3.0$ and $\bar z =3.89$.
}
\label{comparepss}
\end{figure}
The observed flux power spectrum at all three redshift bins is plotted
as the symbols (pentagons at $z=4$, squares at $z=3$ and triangles at
$z=2$) in Figure \ref{comparepss}. Error bars are shown at $z=4$ only to
avoid cluttering. The lines are analytic fits to the data points using
equation (\ref{pspecformula}). These analytic fits will be used in
detail below to measure the amplitude of the power spectrum at each
redshift, but here we use them only to qualitatively visualize the
redshift evolution. The flux power spectrum increases with redshift,
contrary to the expected decrease of the mass power spectrum. The
reason is the fast increase in the mean flux decrement with redshift.
We can understand the effect of the mean flux decrement by considering
the limiting case where all absorbers are optically thin; clearly, for
fixed density fluctuations the amplitude of the flux fluctuations
should grow proportionally to the flux decrement.

\begin{figure}
\plotone{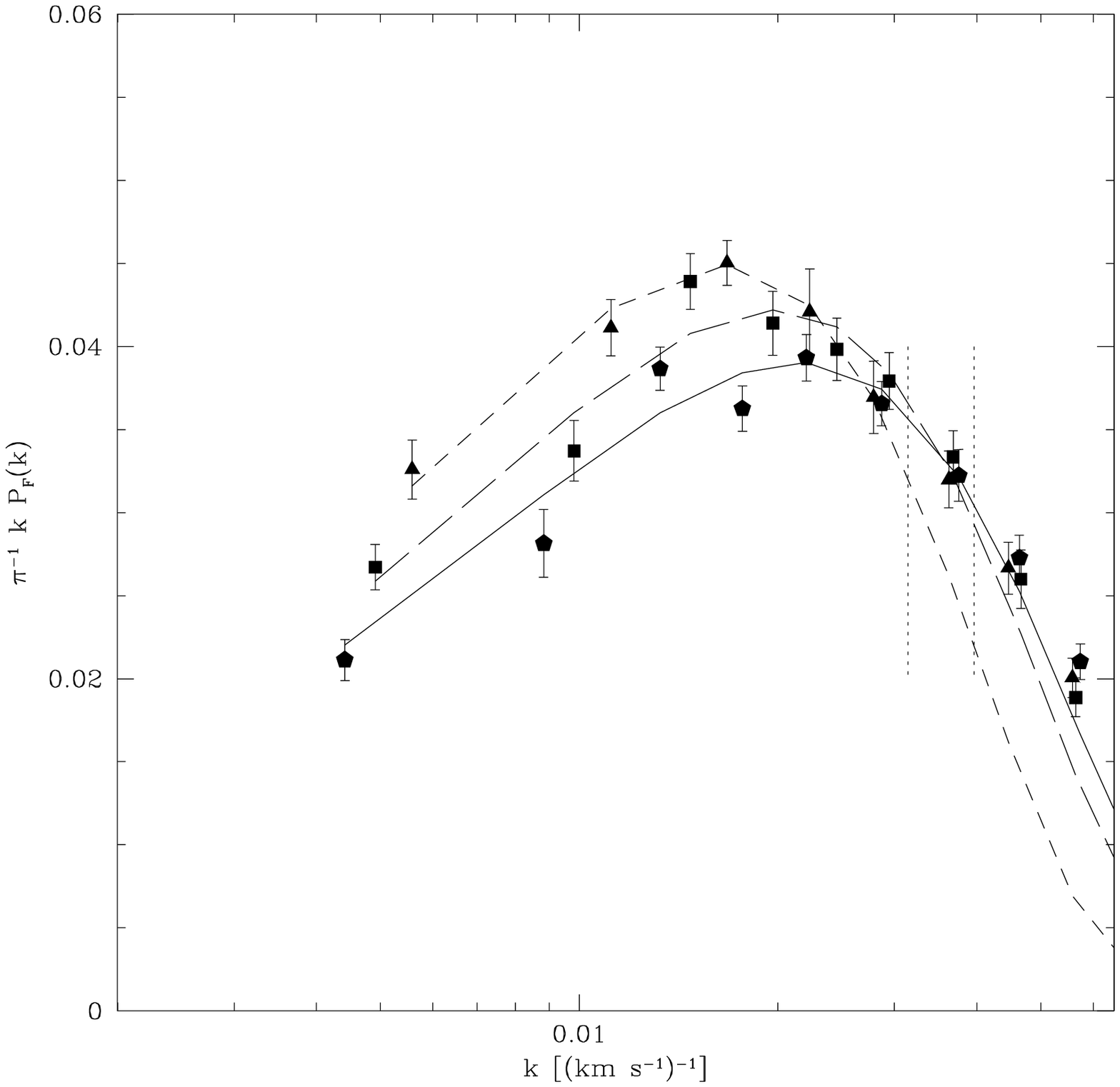}
\caption{The simulated flux power spectrum for the three redshift 
outputs with
optical depths scaled to produce $\bar{F}=0.684$.  The points show the 
values of $P_F(k)$ at the modes of the periodic simulated box, and the
lines are analytic fits to the points.
The redshifts $\bar{z}=4$, 3, and 2 are symbolized by 
pentagons and the solid line, squares and the dotted line, and 
triangles and the dashed line, respectively.  
We see that the large scale power increases with time as expected when
the mean flux decrement is fixed.
The fits use only points to the left of the leftmost vertical dotted
lines for $z=2$, and the rightmost vertical dotted line for the other
two redshifts. Note that the velocity scale corresponding to a given
comoving scale is different for the three redshifts.}
\label{simevps}
\end{figure}

  To see that the power spectrum decreases with redshift as expected
when the effect of the flux decrement evolution is removed, we plot
in Figure \ref{simevps} the flux power spectrum of the numerical
simulation at the redshift outputs $z=2, 3, 4$, normalizing the optical
depth to yield a fixed mean flux decrement $\bar{F}=0.684$ at all
three redshifts. In the limit of large scales, the power is now
increasing as expected. But if the simulated spectra are normalized
instead to the observed flux decrements at $\bar z = 2.41, 3.00$ and
$3.89$, then they have an inverted evolution matching the observed
one, as was shown in Figures \ref{ps3}a,b,c.

  We notice that the flux power spectrum still decreases more slowly
than $a^2$ in Figure \ref{simevps}. The reason for this is more
complex: in the limit of large scales, the flux power spectrum should
generally be related to the mass power spectrum by a constant factor
(see Appendix C),
however this factor depends on the amplitude of fluctuations on the
Jeans scale, which changes with time.  The factor also depends more
weakly on the gas density-temperature relation
and other quantities which are changing with redshift.

\subsection{The Amplitude of the Observed Power Spectrum Relative to
the Simulation}

  We have already compared qualitatively the observed flux power
spectrum observations and the numerical simulation in \S 5.1. Now we
want to use this comparison more quantitatively to determine the
amplitude of the primordial density perturbations required to fit the
observations. The most straightforward method to implement this fit
would be to have several simulations with different amplitudes for the
power spectrum. However, since we have only one simulation, we shall
instead use the three redshift outputs at $z=2,3,4$ as being equivalent
to the results of three different models at the same redshift, having
amplitudes of the initial fluctuations in the proportion $3:4:5$. This
assumes a self-similarity in the evolution of the \lya forest, where
reducing the initial amplitude of the power spectrum is equivalent to
reducing the scale factor by the same factor. This self-similarity is
exact for the dark matter (and assuming $\Omega(z) \simeq 1$). For
gas it is broken on small scales by the hydrodynamic effects of the
temperature. We will be neglecting here any changes in the relation
between the flux and the mass power spectra that are due to a difference
in the effects of the gas temperature at different redshifts. Any such
changes should probably be highly dependent on the heating mechanisms
in the IGM, affected by the model of the ionizing sources and of
reionization that is adopted \citep[see][]{mr94,hg97}.

  We start by comparing each one of the three simulation outputs at
redshifts $z_{sim}=2,3,4$ to each of the three redshift bins in the
data, $z_{obs}=2.41, 3.00, 3.89$. For each one of the nine comparisons,
we go through the following steps in order to compute a $\chi^2$
statistic measuring the degree to which the simulation is consistent
with the observations:

1. We set the mean transmitted flux $\bar{F}$ in the simulation to
match the $\bar{F}$ of the observation, by rescaling the optical depth.
This is an important step because, as we have seen in \S 5.3, the
amplitude of the flux power spectrum is highly sensitive to $\bar F$.

2. We fit the parameters $\Delta^2_F(k_p)$ and $v_c$ (defined in \S 5.2)
to the simulated power spectrum using equation (\ref{pspecformula}),
keeping $n$ and $\Gamma$ fixed to the values of the initial power
spectrum assumed in the simulation.

3. We obtain the dimensionless quantity $k\, P_F(k)$ from the fit to
the simulated power spectrum at redshift $z_{sim}$ (expressing k in
units of the Hubble velocity), and transform it to redshift $z_{obs}$
by rescaling $k$ by the factor $[H(z_{sim})/H(z_{obs})]\,
[(1+z_{obs})/(1+z_{sim})]$, in order to compare the power spectra of
the observations and simulations at the same comoving scale.

4. We compute the $\chi^2$ statistic from the difference between the
observed power spectrum and the fit to the simulation at every $k$ bin
of the observational data, up to $k_{max}= 0.032 \kms$ for all redshift
bins, using the error bars of the observations.

\begin{figure}
\plotone{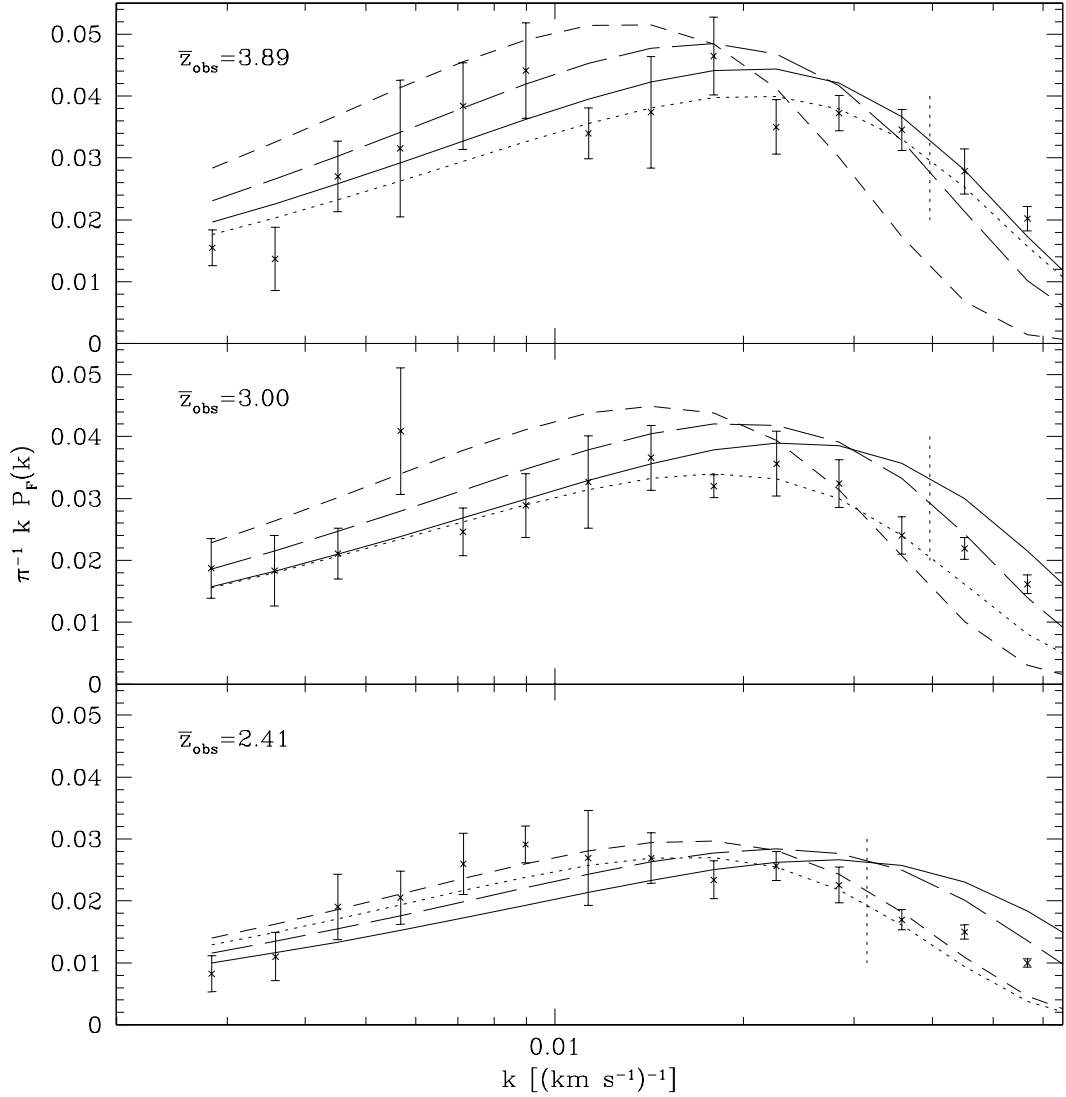}
\caption{The observed power spectrum compared to the fitted simulation
outputs for $z=(4,3,2)$  represented by
({\it solid, long-dashed, dashed}) lines.  The simulated power spectra
are computed after $\bF$ has been fixed to match the observation that
they are being compared to. 
The dotted line is a fit to the observational data. 
}
\label{ps3redscomparesim}
\end{figure}
  In Figure \ref{ps3redscomparesim} we compare the fits to the power
spectra from all three simulation outputs to the observed $P_F(k)$ in
all three redshift bins.  
\begin{deluxetable}{lccc}
\tablecaption{Values of $\chi^2/\nu$ for direct
comparisons between the observational data points and fits to power 
spectrum in the simulation. \label{chisqtab}}
\tablehead{
\colhead{$z_{obs}$} & \colhead{$z_{sim}=4$} & \colhead{$z_{sim}=3$} & 
\colhead{$z_{sim}=2$}}
\startdata
3.89 & 1.8 & 3.6 & 7.6 \\
3.00 & 1.7 & 4.1 & 7.2 \\
2.41 & 2.3 & 1.8 & 1.6 \\
\enddata
\tablecomments{ Each comparison uses only points with
$k<0.032~(\kms)^{-1}$ and has $\nu=11$.}
\end{deluxetable} 
The values of $\chi^2/\nu$ for all of these comparisons (where $\nu$ is
the number of degrees of freedom) are listed in Table \ref{chisqtab}.
It is clear from Figure \ref{ps3redscomparesim} that the amplitude of
the power spectrum needs to be decreased in order to match the
observations at large scales. From Table \ref{chisqtab}, we can see,
for example, that the observations at $z_{obs}=3$ are consistent with
the simulation at $z_{sim}=4$, but inconsistent at $z_{sim}=3$; this
yields an upper bound of $\sigma_8 < 0.79$ for the power spectrum
normalization of the $\Lambda$CDM model of the simulation, since the
amplitude has to be lower than that used in the simulation. At the same
time, the observation at $z_{obs}=2.41$ is significantly better fit
by the simulation output at $z_{sim}=3$ than the one at $z_{sim}=4$;
to within the significance level implied by the difference in the
$\chi^2$ in Table \ref{chisqtab},
this implies a lower limit $\sigma_8 > 0.54$.

  Because the amplitudes of the simulation outputs appear to bracket the
observations, the obvious next step is to interpolate between the
simulation outputs to obtain a best fit value for the power spectrum
amplitude. For
this purpose, we fit the power-law $\Delta_F^2(k_p) = C\,
a_{sim}^{\alpha}$ to the three values of the amplitude $\Delta_F^2(k_p)$
at $z_{sim}=2,3,4$ (where $a_{sim}\equiv (1+z_{sim})^{-1}$) listed in
Table \ref{fittab}, which were obtained by fitting equation
(\ref{pspecformula}) to the simulated power spectrum in \S 5.2. We do
this for all three values of the mean transmitted flux at the three
redshift bins of the data.
\begin{deluxetable}{cccc}
\tablecaption{Parameters for interpolation between the power spectrum 
amplitudes at different redshifts in the simulation.
\label{extraptab}}
\tablehead{
\colhead{$\bar{F}$} & \colhead{$C(\bar{F})$} & 
\colhead{$\alpha(\bar{F})$}} 
\startdata
0.475 & 0.0513 & 0.735 \\
0.684 & 0.0440 & 0.74 \\
0.818 & 0.0288 & 0.670 \\ 
\enddata
\tablecomments{ 
The interpolation formula is 
$\Delta_F^2(k_p)=
C(\bar{F}) (a_{sim}/0.25)^{\alpha(\bar{F})}$.}
\end{deluxetable}  
The results are given in Table \ref{extraptab}.

  To find the best fit for the power spectrum amplitude, we now
fit the observational data at each redshift bin with the two
parameters $\Delta_F^2(k_p)$ and $v_c$, but expressing the
amplitude of the flux power spectrum in terms of $a_{obs}/a_{sim}$,
where $a_{sim}$ is determined from the power-law fit $\Delta_F^2(k_p) =
C \, a_{sim}^{\alpha}$ with the parameters in Table \ref{extraptab}.
\begin{figure}
\plotone{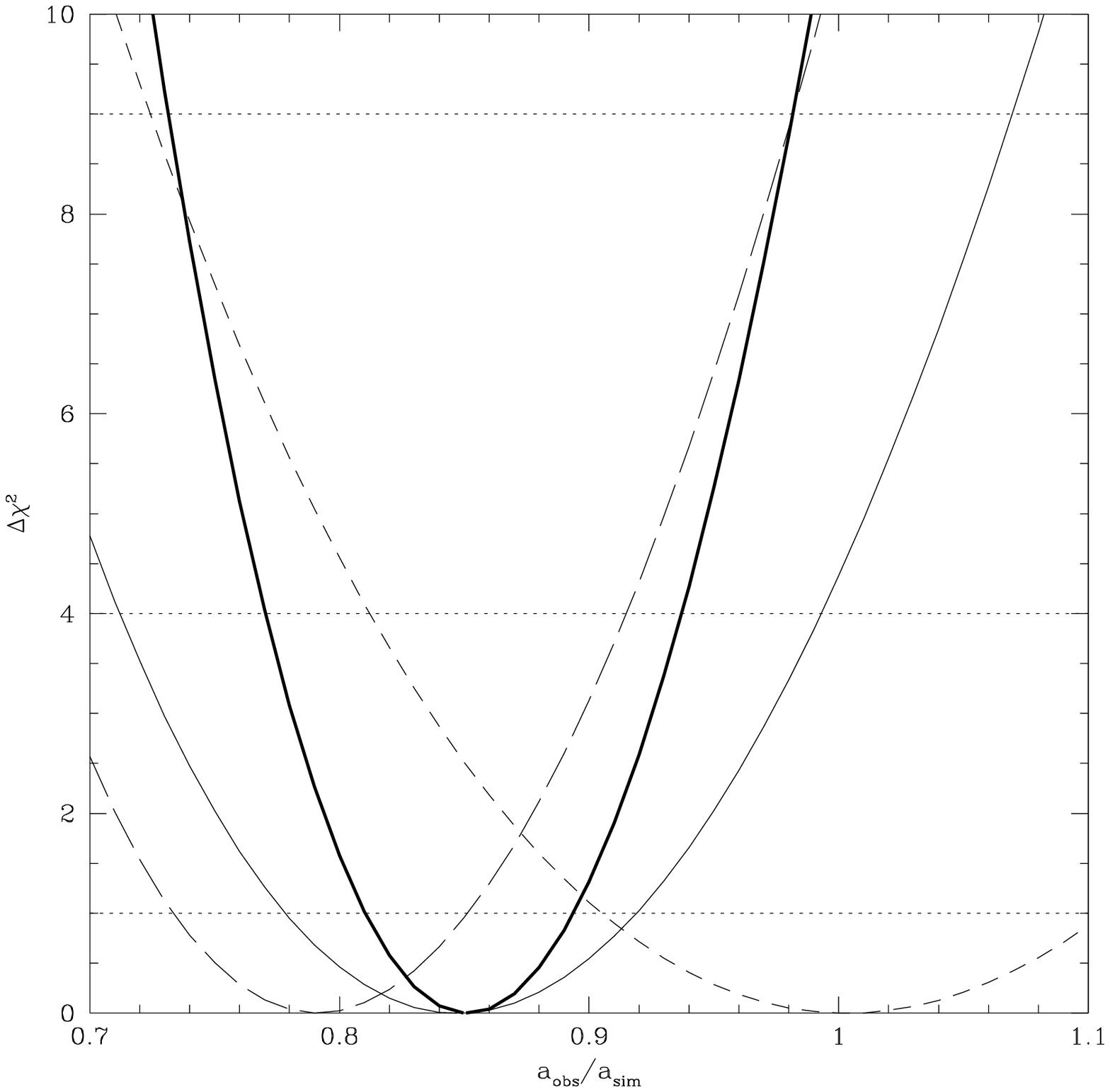}
\caption{Fits comparing the interpolated amplitude in the simulation to
the amplitude of the observations. The thin lines show $\Delta \chi^2$
for $z_{obs}=(3.89,3.00,2.41)$ ({\it solid, long-dashed, short-dashed}).
The thick line shows the combination of all three redshifts.}
\label{chisqAmp}
\end{figure}
Figure \ref{chisqAmp} shows the value of $\Delta\chi^2$ as a function of
$a_{obs}/a_{sim}$, when we let $v_c$ vary, for each redshift bin
$z_{obs}$ ($\Delta\chi^2$ is equal to the $\chi^2$ function minus its
minimum value). The solid thick line is the sum of all three $\chi^2$
functions (subtracting the minimum value), thereby yielding the value and
error of the amplitude when it is required to be the same at all three
redshift bins. The result is $a_{obs}/a_{sim}= 0.856\pm 0.042$; this is
equal to the factor by which we need to multiply the initial mass
fluctuations in the simulation to obtain the best match to the observations.
The error bar is increased to $a_{obs}/a_{sim}= 0.856\pm 0.052$ when
including the error in the determination of the power spectrum
amplitude from the simulation (see Table \ref{fittab}, we
use the error on the $z=3$, $\bar{F}=0.684$ amplitude), determined as 
explained in \S 5.2. The $\chi^2$ value of the power spectrum fit
of all three redshift bins combined, $\chi^2=35$ with 31 degrees of
freedom (35 data points minus four free parameters), should be exceeded
randomly 28\% of the time. This implies that our observational data are
perfectly consistent with the redshift evolution of the flux
power spectrum predicted by the simulation, and with the shape in
equation (\ref{pspecformula}) when $v_c$ is allowed to vary.

  To conclude, our result is that the 
rms amplitude of the initial fluctuations of
the $\Lambda$CDM model in the simulation we use should be reduced by
the factor $0.856\pm 0.052$, or to $\sigma_8 = 0.68 \pm 0.04$.
A possible modeling error in the determination of this amplitude is
that the flux power spectrum is also affected by the temperature
distribution of the gas, and we are relying on the temperatures given by
the simulation for our measurement of the amplitude. For example, if
a temperature-density relation $T\propto \rho^{\gamma-1}$ is followed, the
neutral density varies with density as 
$n \propto \rho^{2-0.7(\gamma-1)}$
in photoionization equilibrium, so the relation between the flux and
mass power spectra depends on $\gamma-1$.

  We have assumed in our analysis that the error bars on the measured
power spectrum points are uncorrelated. We have verified this by
re-running the amplitude fitting procedure described above on 100 
bootstrap realizations of the observed power spectrum for each redshift
bin. We measure the dispersion in the best fit values of
$a_{obs}/a_{sim}$ and find ($\pm 0.054$, $\pm 0.076$, $\pm 0.168$) for
$z=(4, 3, 2)$. The errors estimated using $\Delta \chi^2=1$ in Figure
\ref{chisqAmp} are ($\pm 0.07$, $\pm 0.06$, $\pm 0.11$). The errors at
lower redshift are probably becoming more correlated, as expected given
the increased non-linearity, but the effect is still not large.
We compute a weighted mean and error for $a_{obs}/a_{sim}$ using
these error bars and find 
$a_{obs}/a_{sim}=0.845\pm 0.043$ while we found 
$a_{obs}/a_{sim}=0.856\pm 0.042$ above.  The difference is clearly
not important for our current level of precision, but it may become
important when a larger sample of quasars is available.

\subsection{The Slope of the Power Spectrum}\label{slopenovaramp}

  We now show that the slope parameter $n$ of the power spectrum we
measure from both the simulation and the observations (by fitting
equation [\ref{pspecformula}]) is consistent with the linear theory,
real space power spectrum used to set the initial conditions of the
simulation.
\begin{figure}
\plotone{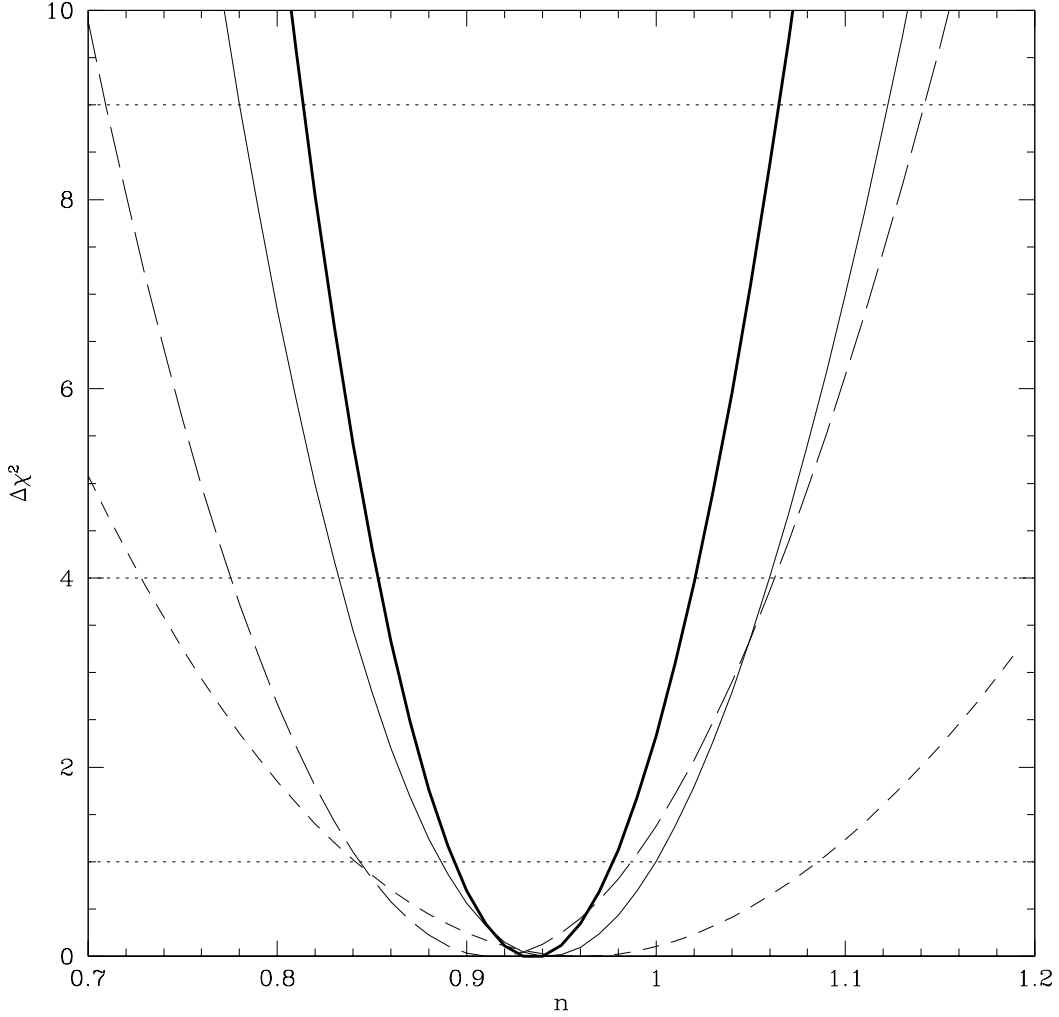}
\caption{Fits to the power spectrum in the simulation with varying $n$.
The linear theory value of $n$ in the initial conditions was $n=0.95$.
The thin lines are $\Delta \chi^2=\chi^2-\chi^2_{min}$ for $z=(4,3,2)$
({\it solid, long-dashed, short-dashed lines}).
The thick solid line is the sum of the three redshifts.
At each redshift the mean flux decrement was fixed to the value of the
nearest observational redshift bin.  
For each redshift the amplitude and a 1D cutoff were fitted 
independently.  Note that the statistical error on the combined result
is larger than $\Delta \chi^2$ indicates because the three sets of
errors are not independent.  The best fit value is $n=0.93\pm0.07$
where the error $\pm 0.07$ is the error for the $z=3$ output alone, 
{\it not} the sum of the three.
The best fits have $\chi^2/\nu = (2.7,1.6,0.16)$ for $z=(4,3,2)$ and
$\nu=(4,4,2)$.  The probability for the $z=4$ value of $\chi^2$ is 
only $3\%$ but this is probably reasonable because we expect that the
error bars on the simulation are somewhat underestimated (see text).
}
\label{slopechisqsim}
\end{figure}

We start analyzing the spectra of the numerical simulation.
In Figure \ref{slopechisqsim}, the $\Delta \chi^2$ value is plotted as
a function of $n$, where $A_F$ and $v_c$ are free parameters
at each redshift. The parameter $\Gamma(z)$ in the power spectrum
formula is still fixed to the model assumed in the simulation.
The result is consistent with $n=0.95$ at all three redshift bins.
The best fit is $n=0.93\pm 0.07$.  We use the
error bar given by $\Delta \chi^2=1$ {\it for the $z=3$ output only}
because the three curves are largely evolved versions of each
other, so the values of $n$ obtained from different simulation outputs
are not independent.
It is especially reassuring to see that the widely different
mean flux decrements at the different redshifts do not seem to have
any systematic effect on the slope.  

\begin{figure}
\plotone{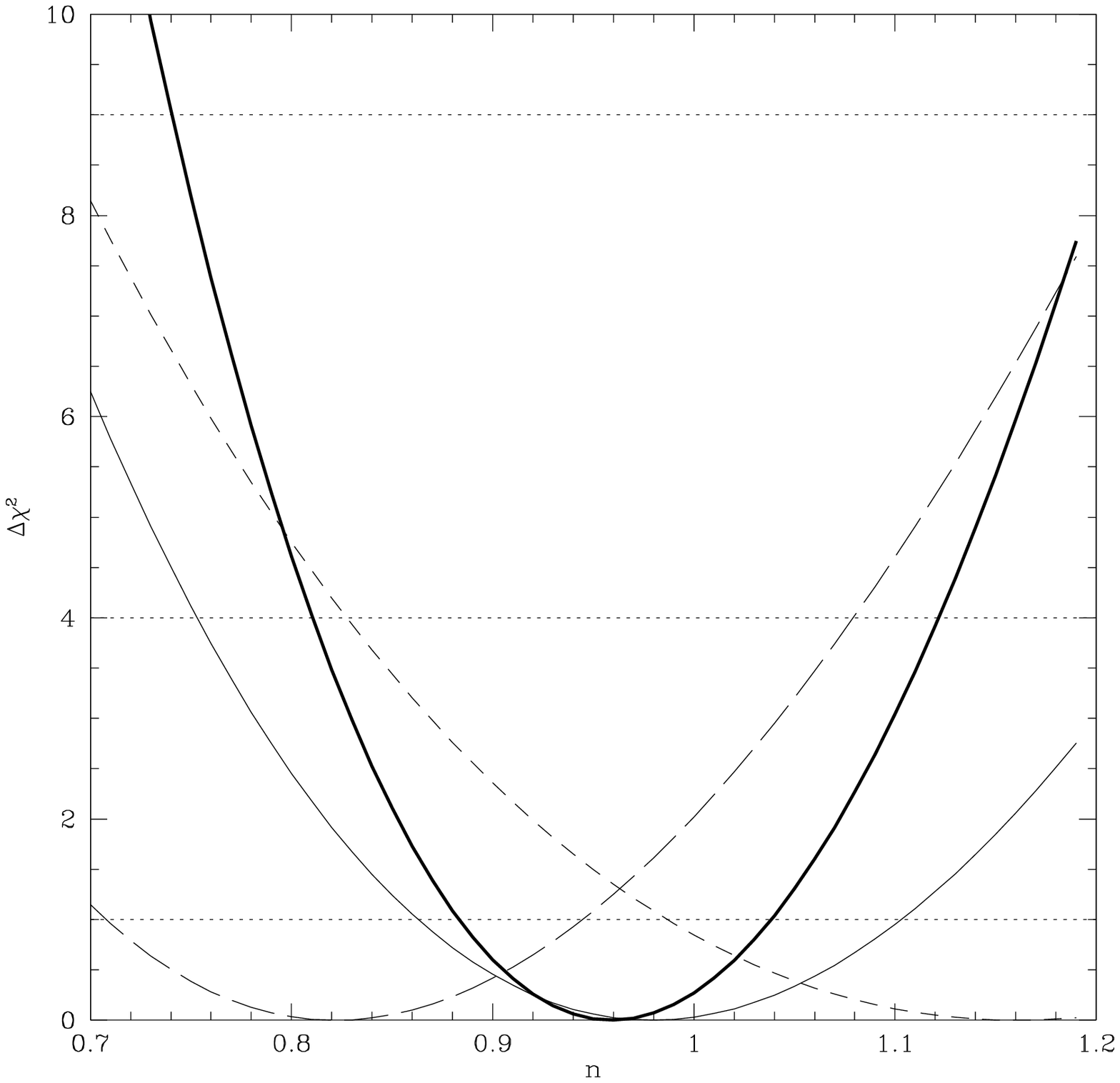}
\caption{Fits to the power spectrum of the observations with varying 
$n$.
The thin lines are $\Delta \chi^2=\chi^2-\chi^2_{min}$ for 
$z=(3.81,3.0,2.41)$
({\it solid, long-dashed, short-dashed lines}).
The thick solid line is the sum of the three redshifts.  
For each redshift the amplitude and a 1D cutoff were fitted 
independently. The best fit value overall is $n=0.96\pm0.08$
with $\chi^2/\nu = 1.13$ for $\nu=28$.
The simulation measurement should be used as a correction to give
a final result $n=0.98\pm 0.11$.   }
\label{slopechisqreal}
\end{figure}
In Figure \ref{slopechisqreal} we present the same $\Delta \chi^2$ 
functions for the observations. We again allow $A_F$ and $v_c$ to 
vary independently at each redshift (six free parameters) but fix
$\Gamma(z)$ to the value given by the 
simulation parameters.
We find an overall best fit value of $n=0.96 \pm 0.08$ with 
$\chi^2/\nu=1.13$ for $\nu=28$ degrees of freedom.

  Since our test of the slope fitting procedure using the simulation
(Figure \ref{slopechisqsim}) yielded $n=0.93\pm 0.07$, when the true
linear theory slope of the model was $n=0.95$, we shall assume that the
same offset $\Delta n = -0.02\pm 0.07$ should be present in the observations.
We use this offset as a correction to the observational slope and add the
error in quadrature. The final result for the estimated
linear theory slope is then $n=0.98\pm 0.11$.
Note that this result assumes that the deviation from a power-law for
the power spectrum of the real 
universe matches the deviation in the $\Lambda CDM$ model of our
simulation.

\subsection{Joint Fit for the Amplitude and Slope}

  The result for the slope in \S 5.5 was obtained independently in the
three redshift bins, without requiring the amplitudes to follow the
evolutionary law of Table \ref{extraptab}. In fact, generally it would
not be valid to use the values of $C(\bF)$ and $\alpha(\bF)$ in this
Table when the initial power spectrum does not have the same slope as
in the simulation. However, the result we found for the slope is very
close to that of the simulation. This justifies a joint fit for the
amplitude and slope using the interpolation constants in Table
\ref{extraptab}.

  The joint fit applied to the simulation yields $a_{obs}/a_{sim}=1$, by
definition, and $n=0.94\pm 0.04$. This result is close to that found in
\S 5.5, with a smaller error bar mostly because this is now a combined
fit to all three redshifts. For the same fit to the observations, the
amplitude is $a_{obs}/a_{sim}=0.85\pm 0.04$ and the slope is
$n=0.92\pm 0.07$. The reduction by two in the number of free parameters
(since the amplitudes at the three redshifts are now tied by the formula
from Table \ref{extraptab}) is accompanied by an increase of 3.3 in
$\chi^2$ for the best fit to the observations. Including the correction
and error from the fit to the slope in the simulation ($\Delta n =
-0.01 \pm 0.07$ using the error bar from only one redshift output) 
gives a final result $n=0.93\pm 0.10$.  

  The value of $k$ we use to specify the amplitude, 
$k_p = 0.04~(\kms)^{-1}$ at $z=3$, was chosen to make the error bars on the slope and
amplitude independent for this measurement. Note that this pivot $k$ is
five times larger than the pivot used by \citet{cwp99} because we have
used data on smaller scales and 
also weighted the fits using the statistical error bars instead of
the more conservative weighting (that favored large scales)
used by \citet{cwp99}.  
The fact that we use the 1D power spectrum
for fitting also contributes to the large value of our $k_p$, because
a data point $P_{1D}(k)$ is an integral over the 3D power 
at wave numbers larger than $k$. 
We are able to take advantage of our higher resolution data because 
we also have a higher resolution simulation 
to test the procedure. However, these errors are statistical only, and
do not include systematic errors due to the way we model the flux power
spectrum as a function of the mass power spectrum based on the
simulation used here.

  Our result is that the observational data 
appears to favor a value of $\sqrt{P(k_p)}$ at $k_p=0.04~(\kms)^{-1}$
that is $0.85\pm~0.05$ times the amplitude of the simulation (after
adding the error in the determination of the simulation amplitude).
Computing the value of the linear theory power in the simulation at
$k_p=0.04~(\kms)^{-1}$ and $z=3$ we find that our result corresponds
to $P_{3D}[k=0.04~(\kms)^{-1},~z=3]=
(2.2\pm 0.3)\times 10^5~(\kms)^3$ or
equivalently
$\Delta_\rho^2[k=0.04~(\kms)^{-1},~z=3]=0.72\pm 0.09$ where 
$\Delta_\rho^2(k)$ is the contribution to the mass density 
variance per unit interval in $\ln k$.
\citet{cwp99} measured
$\Delta_\rho^2[k=0.008~(\kms)^{-1},~z=2.5]=0.57^{+0.26}_{-0.18}$.
Using our amplitude measurement at $k_p$ and our measurement of the
slope relative to the simulation we find 
$\Delta_\rho^2[k=0.008~(\kms)^{-1},~z=2.5]=0.32\pm 0.07$.
The two results are consistent, differing by $\sim 1.3~\sigma$.
Note that the change in slope in the CDM power spectrum, between
the two pivot points, is important for this comparison.

  We have discussed the slope of the power spectrum in terms of the
large scale asymptotic slope $n$ because it is simple to relate this
number to the simulation initial conditions and cosmology in general.
Of course we are really measuring an effective value of the slope on the
scale of the \lya forest. We define the slope parameter $n_p$ to be the
slope at $k_p=0.04~(\kms)^{-1}$ at $z=3$. Using equation
(\ref{pspecformula3D}), it is straightforward to compute $n_p$ for a
given $n$. For the value of $\Gamma(z)$ used in the fits, and with 
$n=0.95$, we find $n_p=-2.53$. The observational result, $n=0.93\pm
0.10$ (or $n=0.98\pm 0.11$ using the more conservative method in \S
5.5), corresponds to $n_p=-2.55\pm 0.10$ (or $n_p=-2.50 \pm 0.11$). For
the pivot point that \citet{cwp99} used, $z=2.5$ and
$k=0.008~(\kms)^{-1}$, the slope of our fitting function is -2.22 for 
$n=0.95$ so our result corresponds to -2.24 (-2.19) while they found
-2.25. For measuring the slope at $k_p$, using the CDM power spectrum
shape is not crucially important; however, the extrapolation we use to
compare to the \citet{cwp99} result is only valid if the CDM power
spectrum correctly describes the shape of the power spectrum in the real
universe. 

\subsection{The $\Delta^2_F-\Delta^2_{mass}$ Relation and Possible
Modeling Errors in our Result}

\begin{deluxetable}{lccc}
\tablecaption{Ratio of flux fluctuations to
linearly extrapolated mass fluctuations in simulation,
$B = [\Delta_F^2(k_p)/\Delta_{mass}^2(k_p)]^{1/2}$.
\label{biastab}}
\tablehead{
\colhead{$z_{sim}$} & \colhead{$B(\bF=0.475)$} & 
\colhead{$B(\bF=0.684)$} & \colhead{$B(\bF=0.818)$}} 
\startdata
4 & 0.264 & 0.243 & 0.198 \\
3 & 0.228 & 0.212 & 0.172 \\
2 & 0.191 & 0.177 & 0.142 \\
\enddata
\end{deluxetable} 
  The ratio of the flux fluctuation amplitude (from Table \ref{fittab})
to the linearly extrapolated mass fluctuation amplitude, $B \equiv
[\Delta_F^2/\Delta_{mass}^2]^{1/2}$, is given in Table \ref{biastab}.
This ratio can be thought of as the ``bias'' of the \lya forest.
However, this factor is not the bias to be applied in linear theory
peculiar velocity calculations, because in the \lya forest
a non-linear mapping is applied to the {\it redshift space} optical
depth field to obtain the flux.

  The results in Table \ref{biastab} should allow for an easy
comparison with other numerical simulations of the quantity we need to
obtain the mass power spectrum amplitude from observations. One of the
main uncertainties in the $B$ factor arises from the
temperature-density relation, which affects the relation between
neutral density and baryon density: $n_{\hi} \propto \alpha(T)~\Db^2
\propto \Db^{2-0.7~(\gmo)}$, where $\Db$ is the baryon density divided
by its mean value, and we have assumed $T\propto \Db^{\gmo}$.
Increasing $\gamma$ decreases the optical depth fluctuations for fixed
baryon density fluctuations. The theoretical expectation is $0.0 < \gmo
< 0.6$ \citep{hg97}, and our simulation falls roughly in the center of
this range, giving a potential error of $\sim 10$\% in $2-0.7~(\gmo)$.
A change in thermal broadening also changes the amplitude of flux
fluctuations in a more complicated way by changing the weighting of
different densities in the spectra. 

\section{THE CORRELATION FUNCTION}\label{corfuncsection}

  We now discuss the correlation function $\xi(\Delta v)=
\left< \delta F(v) \delta F(v+\Delta v) \right>$.  
\begin{deluxetable}{llccc}
\tablecolumns{5}
\tablecaption{The observed flux correlation function.
\label{xitab}}
\tablehead{
\colhead{$\Delta v_{bin,min}$} & \colhead{$\Delta v_{mean}$} & 
\colhead{$\xi(\Delta v,z=3.89)$} & \colhead{$\xi(\Delta v,z=3.0)$} & 
\colhead{$\xi(\Delta v,z=2.41)$} \\
\colhead{($\kms$)}&\colhead{($\kms$)}&&&}
\startdata
7     & 10.5  & $0.1230\pm0.0031$ & $0.1132\pm0.0055$ & $0.0767\pm0.0069$ \\ 
14    & 17.5  & $0.1142\pm0.0032$ & $0.1071\pm0.0055$ & $0.0721\pm0.0067$ \\ 
21    & 24.5  & $0.1039\pm0.0035$ & $0.0993\pm0.0055$ & $0.0667\pm0.0065$ \\ 
28    & 31.5  & $0.0936\pm0.0038$ & $0.0911\pm0.0055$ & $0.0609\pm0.0062$ \\ 
35    & 39.53 & $0.0827\pm0.0040$ & $0.0820\pm0.0056$ & $0.0544\pm0.0059$ \\ 
44.06 & 49.76 & $0.0710\pm0.0043$ & $0.0717\pm0.0057$ & $0.0469\pm0.0056$ \\ 
55.47 & 62.65 & $0.0595\pm0.0046$ & $0.0613\pm0.0059$ & $0.0390\pm0.0051$ \\ 
69.83 & 78.87 & $0.0484\pm0.0048$ & $0.0514\pm0.0061$ & $0.0313\pm0.0047$ \\ 
87.92 & 99.29 & $0.0380\pm0.0049$ & $0.0420\pm0.0062$ & $0.0240\pm0.0041$ \\ 
110.7 & 125   & $0.0294\pm0.0048$ & $0.0336\pm0.0062$ & $0.0178\pm0.0035$ \\ 
139.3 & 157.4 & $0.0211\pm0.0047$ & $0.0275\pm0.0056$ & $0.0120\pm0.0031$ \\ 
175.4 & 198.1 & $0.0143\pm0.0047$ & $0.0227\pm0.0048$ & $0.0069\pm0.0026$ \\ 
220.8 & 249.4 & $0.0095\pm0.0049$ & $0.0177\pm0.0045$ & $0.0037\pm0.0025$ \\ 
278   & 314   & $0.0064\pm0.0046$ & $0.0104\pm0.0041$ & $0.0004\pm0.0023$ \\ 
350   & 395.3 & $0.0047\pm0.0048$ & $0.0078\pm0.0037$ & $-0.0023\pm0.0020$ \\ 
440.6 & 497.6 & $-0.0010\pm0.0055$ & $0.0072\pm0.0035$ & $-0.0003\pm0.0024$ \\ 
554.7 & 626.4 & $0.0027\pm0.0058$ & $0.0029\pm0.0035$ & $0.0014\pm0.0028$ \\ 
698.3 & 788.5 & $0.0060\pm0.0054$ & $0.0033\pm0.0035$ & $0.0025\pm0.0017$ \\ 
879.2 & 992.7 & $0.0059\pm0.0038$ & $0.0060\pm0.0032$ & $0.0005\pm0.0017$ \\ 
1107  & 1250  & $-0.0007\pm0.0040$ & $0.0051\pm0.0045$ & $0.0008\pm0.0017$ \\ 
1393  & 1573  & $-0.0014\pm0.0043$ & $-0.0060\pm0.0041$ & $-0.0005\pm0.0019$ \\ 
1754  & 1980  & $-0.0042\pm0.0033$ & $-0.0005\pm0.0039$ & $-0.0019\pm0.0016$ \\ 
2208  & 2493  & $0.0013\pm0.0033$ & $-0.0028\pm0.0032$ & $0.0002\pm0.0018$ \\ 
\enddata
\tablecomments{ Averaged over bins 
defined by the velocity separation
$\Delta v_{min}$, with mean velocity separation $\Delta v_{mean}$ 
(the maximum velocity separation for the
final bin is $\Delta v_{max}=2780~\kms$).}
\end{deluxetable} 
It is straightforward to measure $\xi$ by directly averaging over the
pixels in the spectra.  The results are shown in Table \ref{xitab}.  
It is important
to recognize that the error bars we give here are {\it strongly} 
correlated and we include them only to give a qualitative idea how 
large they are. Any statistical comparisons with data should use the
full error covariance matrix, which is available in the website given
earlier.
\begin{figure}
\plotone{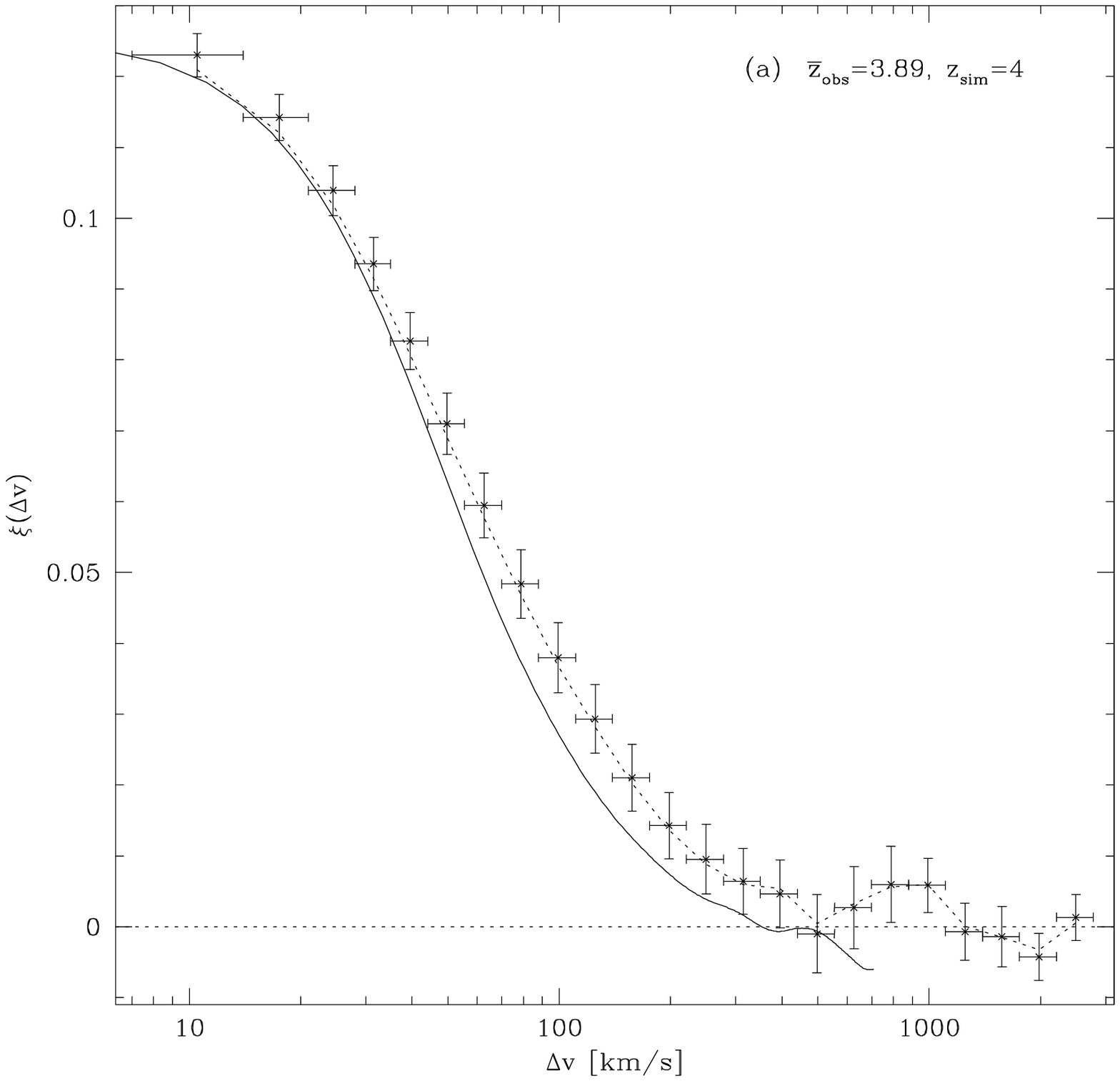}
\caption{The observed correlation function of $F$ as a function of 
velocity 
separation along the line of sight ({\it points with
error bars}).
The definition of the correlation function is:
$\xi(\Delta v)=\left< \delta F(v) \delta F(v+\Delta v) \right>$
where $\delta F = F-\bar{F}$.   The error
bars on $\Delta v$ show the bins used to average the correlation
function.  The error bars on $\xi$ are too highly correlated to use
for statistical analysis without accounting for off-diagonal terms in
the error covariance matrix.
The dotted line shows the correlation function when the possible metal
line regions are included in the computation.
The solid line shows the correlation function from the simulation,
which is strongly influenced by the finite box size. 
(a) shows $\bar{z}=3.89$, (b) shows 
$\bar{z}=3.00$, and (c) shows $\bar{z}=2.41$. }
\label{xi3}
\end{figure}
\begin{figure}
\plotone{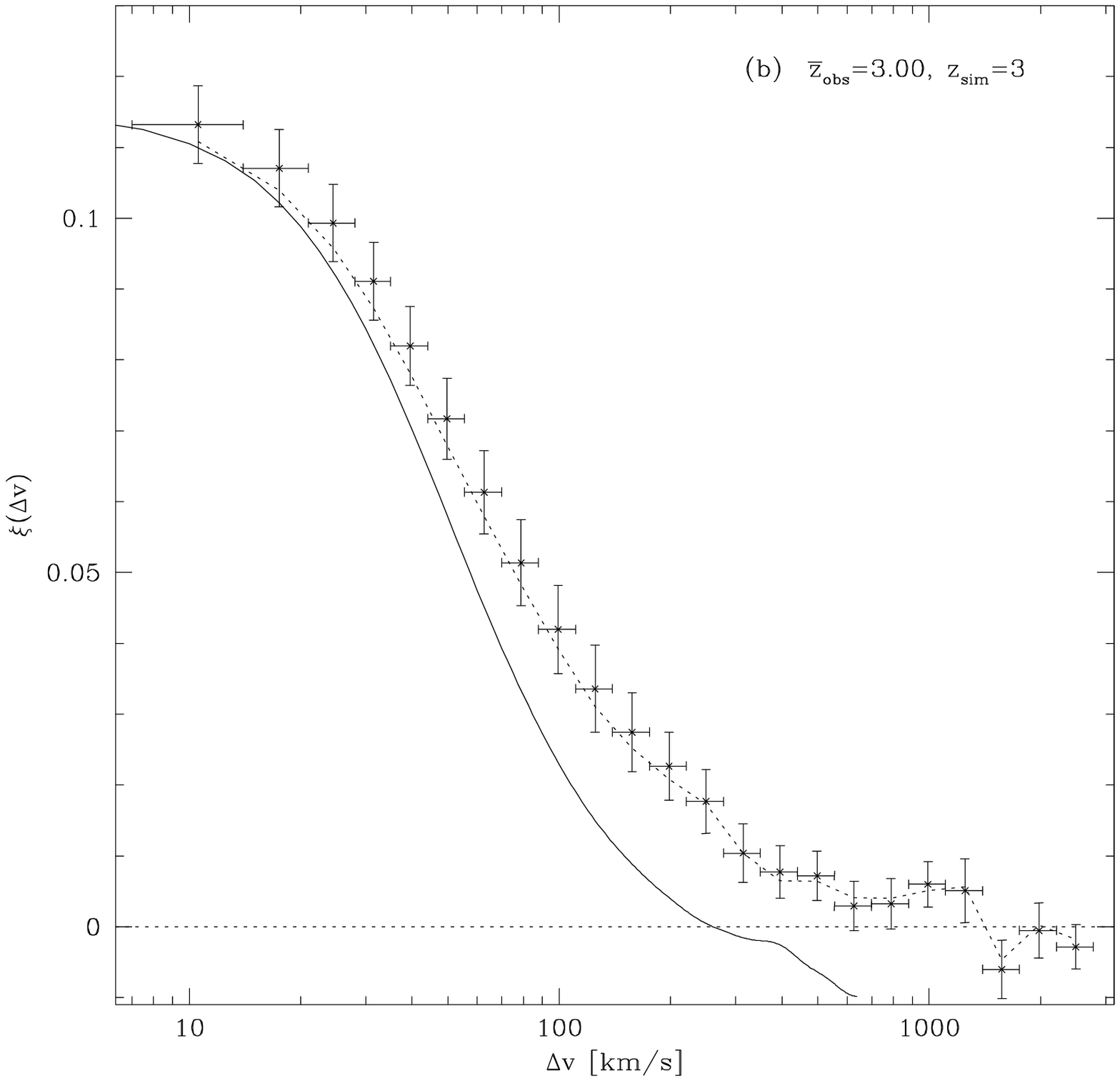}
\end{figure}
\begin{figure}
\plotone{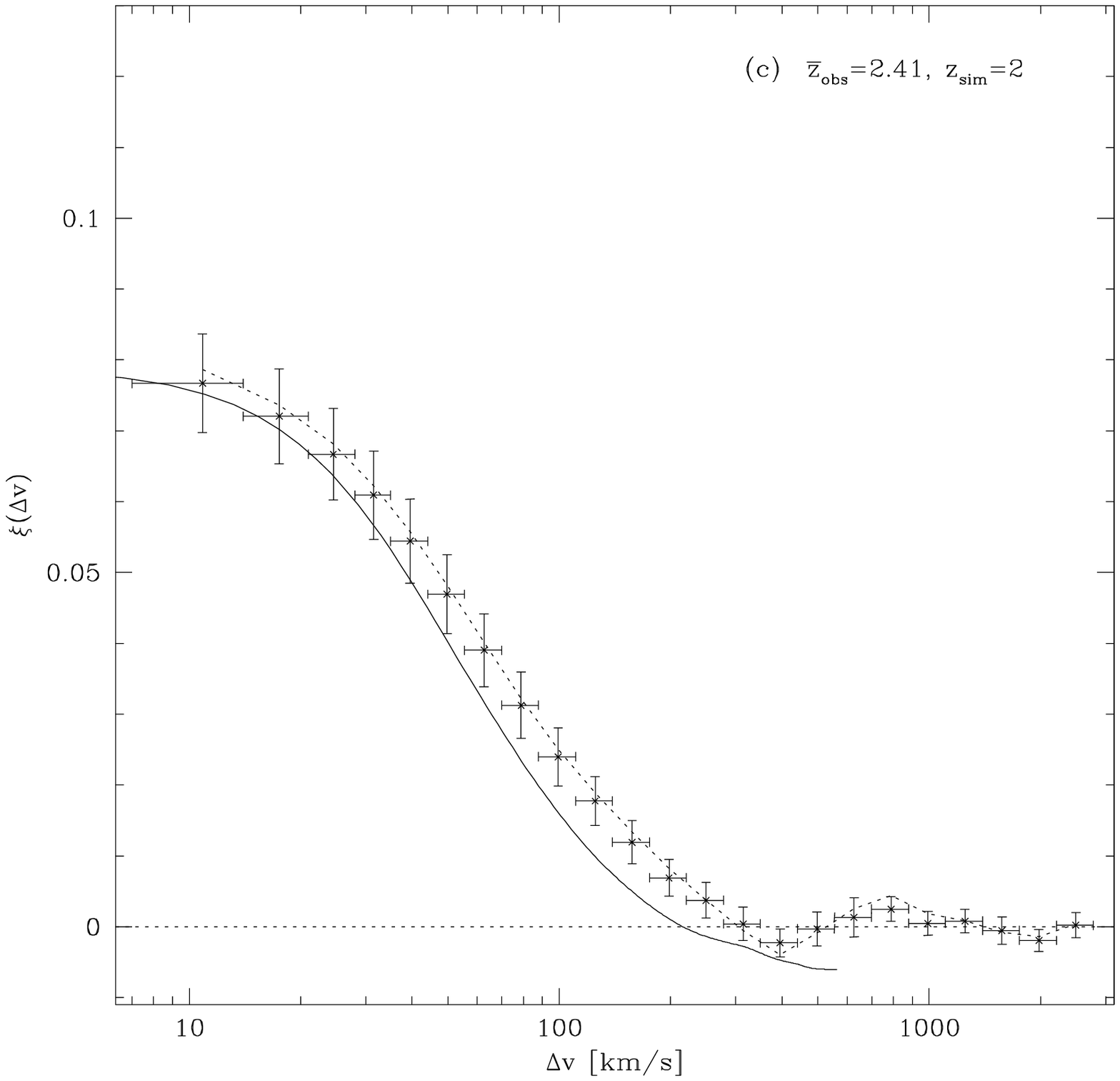}
\end{figure}
In Figures \ref{xi3}(a,b,c) we show the correlation function measured 
from the observations and the simulation.
The observation and simulation do not agree well in spite of the good
agreement between the power spectra shown in Figure \ref{ps3}. The
reason, as we will show below, is that the correlation function computed
from the simulation is strongly affected by the finite box size.
\begin{figure}
\plotone{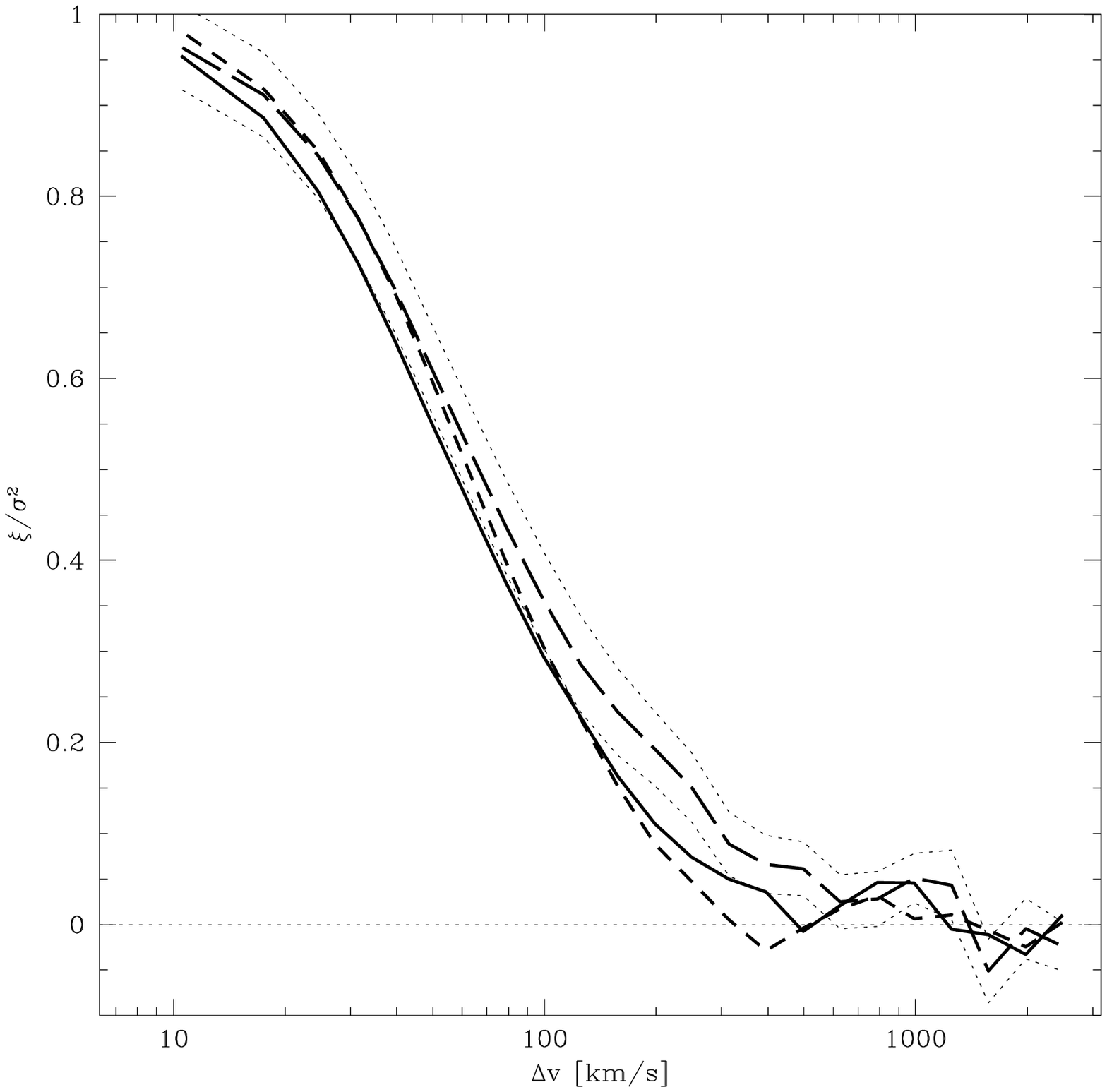}
\caption{Comparison of the correlation functions for $z=3.89$ 
({\it solid line}), $z=3.0$ ({\it long-dashed line}), and 
$z=2.41$ ({\it short-dashed line}).
The amplitude of each curve has been divided by $\sigma_F^2(z)$ so
that $\xi(0)=1.0$ for each.  
The thin dotted lines connect the error bars on the $z=3$ correlation
function.  }
\label{comparexis}
\end{figure}
Figure \ref{comparexis} shows $\xi$ plotted for each of the observed
redshift bins.  For this figure each correlation function has been 
divided by the 
variance, $\sigma_F^2$, at the same redshift, so the curves are all
normalized to $\xi(0)=1$. There is no clear difference between the
correlation lengths at the three redshifts. 
\begin{figure}
\plotone{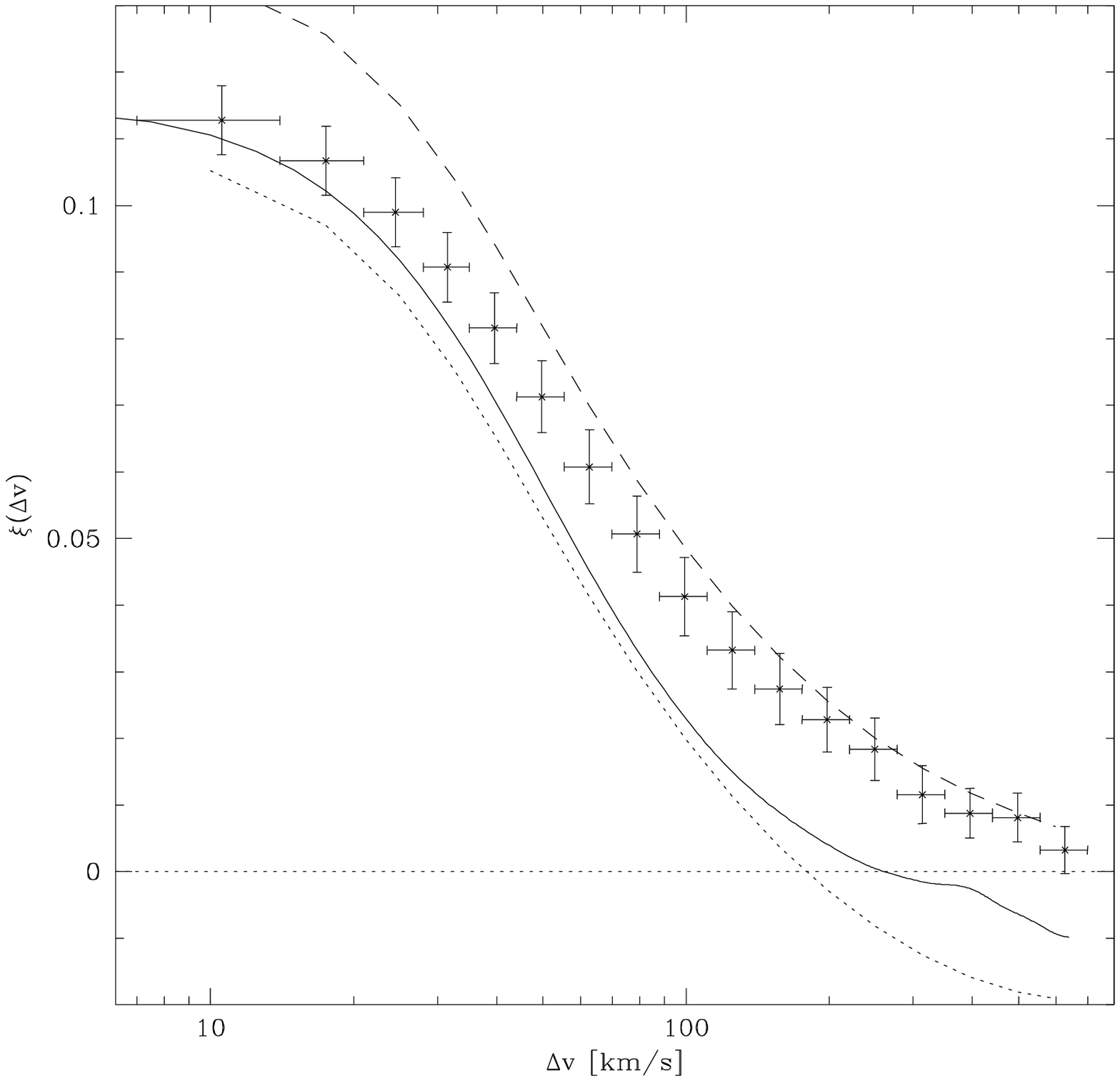}
\caption{The effect of the finite box size of the simulation.  
The points with error bars show the observed $\xi$ for reference.
The solid line shows $\xi$ measured from the simulation.  The dashed
line shows $\xi$ computed by integrating an analytic power spectrum 
with parameters determined by a fit to the power spectrum measured 
from the simulation.  The fit uses a Gaussian
cutoff on the linear theory 3D power spectrum and fits the entire
range of $k$ fairly well.  The dotted line shows the correlation 
function calculated from the same analytic power spectrum, but here
it is computed by summing over the 1D modes actually in the box 
instead 
of integrating over all $k$.  There is a discrepancy between the solid
and dotted lines because the mean along each line of sight through the
simulation can vary from the overall mean and the dotted curve is
computed from a fit to the simulation power spectrum rather than
the power spectrum itself.}
\label{z3xiboxsize}
\end{figure}

Finally we show the effect of the box size on the correlation function
from the simulation in Figure \ref{z3xiboxsize}.  We first fit to the
{\it power spectrum} of the simulation at $z=3$ using the linear
theory function with a 3D cutoff instead of our usual 1D cutoff.  We
use the 3D cutoff because this function fits the full range of $k$, 
including the small scales. When we Fourier transform the analytic
power spectrum using the parameters that fit the simulation, the
correlation function obtained is quite different from the one measured
directly from the simulation. In fact, the agreement in Figure 
\ref{xi3}
between the observed and simulated $\xi(\Delta v)$ as
$\Delta v \rightarrow 0$ is only a coincidence: the observed correlation
function is in fact smaller than in the simulation model, but the
correlation function in the simulation is lowered by the finite box size
effects to a value close to the observed one. Figure \ref{xi3} also shows
that we can approximately recover the correlation function that is
measured from the simulation by summing over the analytic power spectrum
using only $k$ values of discrete modes that are actually included in
the simulation box.  

\section{DISCUSSION}\label{discussion}

  We have presented an observational determination of the PDF, the power
spectrum, and the correlation function of the transmitted flux in the
\lya forest, using a sample of eight quasars over the redshift range
$2.1 < z < 4.4$. All our results are given with error bars that include
the variance of our sample of eight quasars. These results can be
compared to the predictions of any cosmological simulations of a given
theory, after instrumental resolution and noise with the same amplitude
as in our data are added to the simulated spectra. Any future
observational determinations of the same quantities from other data sets
will also be easily comparable.

  In our opinion, there are two main objectives in the accurate
measurement of the statistics of the one-dimensional random field of the
transmitted flux in the \lya forest. The first is to test the idea that
the \lya forest is entirely explained by an IGM that is photoionized and
heated by the known sources of ionizing radiation (AGNs and galaxies),
which evolves gravitationally as described by a large-scale structure
theory based on the presence of cold dark matter and adiabatic, Gaussian
primordial fluctuations. The second objective is to measure some of the
parameters of the large-scale structure model from the \lya forest
observations.

  Even though there are a large number of parameters required to fully
describe the cosmological and large-scale structure model, and the
population of sources of ionizing radiation, only certain combinations
of these parameters are important to determine the observable properties
of the \lya forest to a first approximation. A simple intuitive
understanding of what these parameters are can be developed by
considering a simplified model where the power spectrum of fluctuations
is approximated as a power-law, and the IGM follows a simple
density-temperature relation $T=T_0\, (\rho/\bar\rho)^{\gmo}$. In this
case, the temperature $T_0$ determines a Jeans scale, and all the
statistical properties of the \lya optical depth should depend only on
the amplitude of the density fluctuations at the Jeans scale,
$\sigma_J$, the effective power-law index near the Jeans scale $n_J$,
and $\gamma$ (it is understood here that all these parameters depend on
redshift for any given model). In addition, varying the temperature
$T_0$ (while keeping $\sigma_J$, $n_J$ and $\gamma$ fixed) should cause
a rescaling of all
the statistical properties of the \lya forest in velocity (proportional
to the Jeans scale, or $T_0^{1/2}$), and varying the parameter $\mu$
(defined in eq.\ \ref{mudef}) should cause a rescaling of the optical
depth. Therefore, in this simplified model the \lya forest depends on
five parameters only, and the changes under two of them are trivially
obtained by rescalings. Only $\sigma_J$ and $n_J$ contain
information on the large-scale structure model; $T_0$ and $\gamma$
depend on how the IGM was reionized 
\citep[e.g., ][]{mr94,hg97} as well as the power spectrum (through the
effects of shock-heating), and $\mu$ depends on $\Omega_b$, $H(z)$, and
the intensity of the ionizing background.

  More exact predictions for the \lya forest obtained in hydrodynamic
simulations will depend on more parameters, due to effects that are
neglected in the simplified model just discussed. The true
distribution of density and temperature of the gas has of course a
scatter around a mean relationship; this mean relationship also
deviates from a power-law. This distribution depends on the ratio of
the cooling rate to the Hubble rate, which is proportional to the new
parameter $\Omega_b h^2/H(z)$. In addition, the effects of cooling imply
that changing $T_0$ will no longer result in a simple rescaling in
velocity of all statistical properties of the \lya forest, since the
cooling rate has a complex dependence on temperature that breaks the
self-similar scaling. These effects should be important only at high
densities, where cooling is important; at low densities, the thermal
evolution of the IGM naturally produces the ``Hui-Gnedin relation''
\citep{hg97}. The value of
$\Omega(z)$ can also change the properties of the density field under
gravitational evolution, but this effect should be small because
$\Omega(z)$ is very close to unity at high redshift. The evolution of
$T_0$ with redshift, and the baryon fraction ($\Omega_b/\Omega$) can also
affect the Jeans scale and the gravitational evolution of the
photoionized gas \citep{gh98}. These other parameters are
likely to cause only small changes in observable properties compared
to the five main parameters given above.

  We have presented here measurements of some of the principal
parameters determining the statistical properties of the \lya forest. The
mean transmitted flux determines the parameter $\mu$, and the flux power
spectrum can yield the amplitude and the effective slope of the initial
mass power spectrum. We have carefully computed the error bars due to
sample variance for all our results, which are generally consistent with
previous ones, although the amplitude of the power spectrum is found to
be slightly lower than in \citep{cwp99}. Both the flux PDF and the
power spectrum predicted in the numerical simulations in \citet{mco96} 
agree extremely well with what is observed. The main
limitation we have for the accuracy of this comparison are the
uncertainties introduced by the continuum fitting of the flux, and by the
presence of narrow metal lines.

Our results for the measurement of the initial mass density 
perturbations
are given in terms of the amplitude and power law index (i.e., the
slope on a plot of $\ln P(k)$ vs. $\ln k$) of
the linearly extrapolated primordial power spectrum at 
$k_p=0.04~(\kms)^{-1}$ at $z=3$.  The amplitude is
$\Delta_\rho^2[k=0.04~(\kms)^{-1},~z=3]=0.72\pm 0.09$ where 
$\Delta_\rho^2(k)$ is the contribution to the mass density 
variance per unit interval in $\ln k$.  The slope is
$n_p=-2.55 \pm 0.10$.  This amplitude result, with its small error bar,
must be thought of as partially preliminary.  Further study is needed
to carefully quantify the effects on the $\Delta^2_F-\Delta^2_{mass}$ 
relation of changing the temperature-density
relation or details of the simulations.  \citet{cwk98,cwp99} found no 
important systematic errors but this was in the 
context of larger statistical error bars, and their set of 
simulations was not exhaustive.  

  If it proves to be robust, the power spectrum result can strongly
constrain potential cosmological and galaxy formation models.
For example, \citet{mm99b} found that, in order to produce
the high velocities observed in studies of the kinematics of 
damped \lya systems \citep{pw97,pw98}, a model must
obey $\sigma_{DLA4}>0.78$ (95\% confidence), where $\sigma_{DLA4}$ is 
the linear theory rms fluctuations on spheres of radius $100~\kms$ at
$z=4$. The \lya forest power spectrum amplitude we have derived in this
paper implies $\sigma_{DLA4}=0.67\pm 0.04$, inconsistent with the
above upper limit derived from damped \lya systems at the $2.5 \sigma$
level. This discrepancy, if confirmed in the future, might simply be
due to a different relation between the observed velocity dispersions
in damped \lya systems and the velocity dispersions of their host dark
matter halos from the one that was calculated in the spherical
equilibrium models in \citet{mm99b}, but it should in any case provide
a new constraint to understand the dynamics of damped \lya systems.

  We will also present additional results on the measurement of the
parameters $T_0$ and $\gamma$ in work that is now in preparation
(P. McDonald \etal, in preparation). These
parameters can be determined from the shape of the power spectrum at
small scales, or by fitting absorption features to Voigt profiles and
comparing the distribution of Doppler widths to what is obtained in
simulations \citep{stl99,rgs99,bm99}.

  Apart from measuring these parameters, one should also strive to find
tests for the validity of the general framework of the gravitational
evolution theory in CDM models of the \lya forest. Various perturbations
from this ideal theory should be present at some level. Reionization
smoothes the gas fluctuations at scales smaller than the Jeans scale for
photoionized gas, but dense structures near the Jeans scale will not be
ionized, heated and destroyed for a long time 
\citep{bss88}. For example, a virialized halo with velocity dispersion
$\sim 10 \kms$ should not accrete much gas when forming after
reionization (since the gas has too much initial entropy and is not
able to cool), but having formed before reionization it may keep its
gas in thermal and hydrodynamic equilibrium, as in the minihalo model
\citep{r86,am98}. These objects might introduce
subtle changes in the \lya forest that would depend on when and how the
IGM was reionized. At the same time, when \heii is reionized, the rapid
heating of the IGM can be inhomogeneous on large scales, resulting
in new fluctuations in the \lya forest due to the gas temperature that
depend on the luminosity function and spatial distribution of the
ionizing sources \citep{mr94}. Outflows from AGNs, radio sources and
starbursts may also perturb the IGM at some level. Finally, the
large-scale structure theory itself may be modified (for example, by
having non-Gaussian primordial fluctuations). A continued effort in
comparing the observed spectra with detailed numerical simulations,
using a diversity of statistical probes and methods, is required to
search for the influence of these phenomena.

\acknowledgments
We thank David Weinberg for his useful comments on an early draft of
this paper.

\appendix
\section{GENERATION OF MOCK SPECTRA}

To check various aspects of our data analysis procedure it is useful
to have a method for creating mock spectra with properties similar to
the observed ones.  For example, we test our code for computing power 
spectra from the observations by using it to compute the known 
$P_F(k)$ 
of the mock spectra.
We also test our method for generating error bars by
comparing the bootstrap errors estimated from single random spectra to
the 
ideal results obtained by generating large numbers of independent
random spectra (see below).

  The procedure for generating random
spectra consists of the following steps: \\
\\
1.  Fit an analytic function to the observed $P_F(k)$.\\
2.  Generate a Gaussian random realization of the fitted $P_F(k)$ on 
    a large, fine grid.\\
3.  Map the values from the grid points onto the observed pixel 
    coordinates $\Delta v_i$.     \\
4.  Add noise to each pixel with amplitude given by the mean of the
    observations.   \\ 
\\    
In step 1 we
fit the observed $P_F(k)$ using Equation
(\ref{pspecformula3D}) multiplied by a 3D Gaussian smoothing function 
instead of the 1D smoothing function we generally use.  We use this
function because it provides a good qualitative fit to the observed
$P_F(k)$ over close to the full range of $k$.  We are not concerned 
with
obtaining the best possible fit because we only want to generate mock
spectra with a precisely {\it known} input power spectrum that will 
have properties similar to the observed one.  
In step 2 we generate a Gaussian 
random field with Fourier mode amplitudes given by 
$\left< a_k^2\right>=P_F(k)$ where $P_F(k)$ is given by the fitting 
formula from step 1.  The grid used is evenly spaced in 
velocity and is always longer than the observed
spectrum by a factor of at least 8 with resolution better by a 
factor of at least 4.  
In step 3 we superimpose the uneven set of observed pixel locations 
on top of the grid generated in
step 2 (using a random origin).  
The value of $F_i$ assigned to each $\Delta v_i$ is linearly 
interpolated from the two adjacent values on the fine grid.
The spectra generated by this procedure do not have identical 
statistics to the observations because the observations are not 
Gaussian but this should not be important for our purposes.

  Although the random fields that we generate have the same power 
spectrum as the observed spectra, they have a Gaussian PDF which is not
limited to the range $0<F<1$.  It should be possible to create more
realistic spectra by generating a Gaussian 
(or even non-Gaussian) field for the optical
depth or the baryon density instead of the transmitted flux
\citep[e.g., ][]{b93}.  We have adopted the very simple method of 
generating a Gaussian field and calling it the transmitted flux 
because this allows us to use directly the power spectrum
measured from the observed transmitted flux. 
 
\section{BOOTSTRAP ERROR BARS}

It is well known that
gravitational collapse leads to non-Gaussianity of the density field
in the non-linear regime probed by the smallest scales in the 
\lya forest.
It is generally difficult to estimate error bars on measurements when 
we cannot assume that the underlying process is Gaussian. 
To solve this problem we compute error bars for all of the numbers 
measured in this paper
using a variation of the bootstrap method \citep{ptv92}.

If we have $N$ independent data segments, a
bootstrap realization of a computed statistic is created by randomly 
drawing $N$ segments from the set with replacement and 
recomputing the statistic from the randomly modified set of data.
The bootstrap error bar on a statistic is simply the dispersion 
among the bootstrap realizations of the statistic.
Ideally we would like to divide the data sample into numerous 
independent parts with identical statistical properties as we could
if we had a large number of quasars all at the same 
redshift for example. Unfortunately we only have 8 quasars and 
no more than 5 in a single redshift bin.  One would think that because
the spectra are long relative to the correlation length they can be
split into some number of approximately independent chunks and the 
bootstrap method applied to those chunks.  The question then is how
large must these chunks be relative to the correlation length? 
We can analytically investigate the applicability of the bootstrap 
method for the situation at hand in the special case where we assume 
the field {\it is} Gaussian.  The bootstrap method does not depend on
the field being Gaussian so we can hope that if it works in the
Gaussian case it will work similarly well on observational data with 
the same power spectrum.     

We now investigate analytically the requirements for obtaining 
reliable bootstrap errors in the case where we are segmenting 
correlated data.
We represent an abstract set of data by $\delta_i$ 
where $i=1..N$ runs over all $N$ pixels.  To make analytic calculations
possible we assume $\delta$ is a Gaussian random field with 
$\left<\delta\right>=0$, $\left<\delta^2\right>=\sigma^2=\xi(0)$, and
$\left<\delta_i \delta_j\right>=\xi(i-j)$.  We want to imagine the 
situation where the data has been split into $N/c$ segments each 
containing $c$ pixels.  If $f$ is some statistic that can be 
measured from the data (like the mean), we define $\tilde{f}$ to be 
the value of $f$ that is actually computed from a given set of data,
and $\tilde{\tilde{f}}$ to be the value of $f$ computed from a 
bootstrap realization of the same data used to compute $\tilde{f}$. 
Taking the variance as an example we have
\begin{equation}
\tilde{\sigma^2}=\frac{1}{N}\sum_{i=1}^N \delta_i^2
\end{equation}
and
\begin{equation}
\tilde{\tilde{\sigma^2}}=\frac{1}{N}\sum_{s=0}^{\frac{N}{c}-1}
o_s \sum_{i=s c+1}^{s c+c} \delta_i^2
\end{equation}
where $o_s$ is the number of times that data segment $s$ was drawn
in forming the bootstrap
realization.  The occupation numbers $o_i$ have the following 
property
\begin{equation}
\left<\left(o_i-1\right)\left(o_j-1\right)\right>_b=
\delta_{ij}\left(\frac{N_s}{N_s-1}\right)-\frac{1}{N_s-1}
\end{equation}
where $\left< \right>_b$ represents an average over different 
bootstrap realizations of the same data, $N_s\equiv N/c$ is the 
number of segments drawn from, and $\delta_{ij}=1$ for 
$i=j$ and $\delta_{ij}=0$ for $i\neq j$.
We will generally use the large $N_s$ approximation 
\begin{equation}
\left<\left(o_i-1\right)\left(o_j-1\right)\right>_b=
\delta_{ij}-\frac{1}{N_s} ~.
\end{equation}

To demonstrate the application of this formalism we will compute the
bootstrap error on $\tilde{\sigma^2}$.  To make the algebra more
transparent we will
further simplify the calculation by making the assumption that the
pixels are uncorrelated and the segment length is only one pixel:
\begin{equation}
\left<\left(\tilde{\tilde{\sigma^2}}-\tilde{\sigma^2}\right)^2
\right>_b =
\left<\left(\frac{1}{N}\sum_{i=1}^{N}
(o_i-1) \delta_i^2\right)^2
\right>_b~.
\end{equation}
Evaluating the expectation value over bootstrap realizations gives
\begin{equation}
\left<\left(\tilde{\tilde{\sigma^2}}-\tilde{\sigma^2}\right)^2
\right>_b =
\frac{1}{N^2}\sum_{i=1}^{N} \delta_i^2
\sum_{i'=1}^{N} \delta_{i'}^2
\left(\delta_{i i'}-\frac{1}{N}\right)~.
\label{sigmabs}
\end{equation}
In this simplified case it is easy to find the true error on $\sigma^2$
from the assumption that the field is Gaussian.  The result is
\begin{equation}
\left<\left(\tilde{\sigma^2}-\sigma^2\right)^2
\right>=\frac{2 \sigma^4}{N} ~.
\end{equation}
To relate this to the bootstrap error we must take the expectation 
value of Equation (\ref{sigmabs}) over different realizations of the
underlying Gaussian distribution to obtain
\begin{equation}
\left<\left<\left(\tilde{\tilde{\sigma^2}}-\tilde{\sigma^2}\right)^2
\right>_b\right>=\frac{1}{N^2}\sum_i 3 \sigma^4-
\frac{1}{N^3}\sum_i\sum_j
\left(\sigma^4+2 \sigma^4 \delta_{ij}\right)
=\frac{2 \sigma^4}{N}
\end{equation}
where we have completed all of the sums using $\left<\delta_i \delta_j
\right>=\xi(i-j)=\delta_{ij}$ 
and dropped terms of higher order in $N^{-1}$.  We see that the correct
result is obtained in the large $N$ limit.

The error on the correlation function is more complicated and we will
not assume uncorrelated pixels or one-pixel segments.
The exact result for a Gaussian field in the limit that the data set
is much longer than the correlation length is
\begin{equation}
\left<\left(\tilde{\xi_n}-\xi_n \right)
\left(\tilde{\xi_m}-\xi_m \right)
\right>=\frac{1}{N}\sum_{k=-N}^{N}\left(\xi_k \xi_{k+n-m} +
\xi_{k-m} \xi_{k+n}\right) 
\label{xierror}
\end{equation}
where we have also assumed that $n$ and $m$ are much smaller than the
length of the data set.
A bootstrap realization of $\xi$ is given by
\begin{equation}
\tilde{\tilde{\xi_n}}=\frac{1}{N}\sum_{s=0}^{\frac{N}{c}-1} o_s
\sum_{i=s c +1}^{s c +c} \delta_i \delta_{i+n} ~.
\end{equation}
After evaluating the bootstrap average we find
\begin{equation}
\left<\left(\tilde{\tilde{\xi_n}}-\tilde{\xi_n} \right)
\left(\tilde{\tilde{\xi_m}}-\tilde{\xi_m} \right)
\right>_b=\frac{1}{N^2}\sum_{s=0}^{\frac{N}{c}-1}
\sum_{s'=0}^{\frac{N}{c}-1}\sum_{i=s c +1}^{s c +c}
\sum_{i'=s' c +1}^{s' c +c} \delta_i \delta_{i+n} \delta_j 
\delta_{j+m} \left(\delta_{s s'} -\frac{c}{N}\right) ~.
\end{equation}
If we make the assumption that the data set is long enough that it 
can be split into many segments with each segment much longer than
the correlation length the
calculation of the bootstrap errors averaged over realizations of the
Gaussian field gives the correct result:
\begin{equation}
\left<\left<\left(\tilde{\tilde{\xi_n}}-\tilde{\xi_n} \right)
\left(\tilde{\tilde{\xi_m}}-\tilde{\xi_m} \right)
\right>_b\right>=\frac{1}{N}\sum_{k=-c}^{c}\left(\xi_k \xi_{k+n-m} +
\xi_{k-m} \xi_{k+n}\right) ~.
\end{equation}
The assumptions we were forced to make so that the bootstrap error
came out correct seem to be the ones that would be expected 
intuitively:  the data set must be split into a large number of 
segments and those segments must be larger than the correlation 
length of the data.  Note that cross correlations for widely separated
$n$ and $m$ will not be accurately reproduced but only if they are
so widely separated that the cross correlation is in fact very small. 
We have really only shown that the bootstrap method can
be made to work in the infinite data limit.  

To investigate the effectiveness of the bootstrap method applied to 
real data we turn to the mock spectra described in Appendix A. 
We generate a set of 50 of each quasar spectrum and apply
the data analysis procedure to them exactly as we have done with the
observational data.  We find that the optimal number of segments for
our redshift bins seems to be about 300 giving segments approximately 
100 pixels long.   
The errors on $\sigma_F^2$, the PDF, and $\xi(\Delta v)$ are 
underestimated by 5-15\% for the three redshift bins using this size
segments.  The error estimated for the mean transmitted flux seems to
be more strongly affected by the long range correlations
(underestimated by a factor of 1.6 in the worst case), probably 
because it is only first order in $\xi$ instead of second order like
we see in Equation (\ref{xierror}).

To correct for the these underestimations we find the ratio of the
true error to the bootstrap error in the mock spectra for every 
statistic we measure and then multiply the bootstrap error on the
observational measurement by this factor.  In the case of the PDF
and correlation function we have a full error matrix including the
correlations between errors on different bins.  We judge the 
underestimation of the errors by looking at $\chi^2$ tests  
comparing the PDF or $\xi$ for all possible of pairs of mock
data sets.  The mean value of $\chi^2/\nu$ for $\nu$ degrees of 
freedom should be 1 and the full 
distribution of $\chi^2$ can be computed easily.  We find that the
mean $\chi^2$ computed using the bootstrap errors is no more than
20\% higher than expected and we rescale the error matrix of
the observational data by this overall factor instead of trying to 
correct the individual matrix elements.  With this rescaling the 
distribution of $\chi^2$ becomes essentially perfect.
Note that we do not in any way compute the observational errors from 
the 
Gaussian mock spectra since we use them only to find edge effect 
corrections that are then applied to the bootstrap errors calculated
from the observational data.

Although it plays a large part in the paper, in this discussion we 
have left out any mention of the power spectrum.  We have not found
a fully satisfactory method for computing a full error matrix for
$P_F(k)$ using the bootstrap method.  We do not get correct results 
using the type of analytic calculations we did for the correlation 
function.  This is probably because we need to treat the edge effects
from the finite data set more carefully in the case of the power 
spectrum.  The obvious technical problem with simply applying the
method to the data anyway is that we can not make the segments very
small and still measure the power spectrum from them.  We need to 
keep the segments a few times larger than the longest wavelength for
which we want to measure $P(k)$ but this restricts the number of them
for our
data set to about 10.  We have tried computing full error matrices
for our mock spectra using only 10 segments but the results are 
discouraging.  The distribution of $\chi^2$ computed from these 
matrices is not close to what is expected.  This is because the 
statistical error in estimating all of the error matrix elements from
only ten segments is too high.  Note that  
in our mock spectra there are {\it no} correlations between
the $P_F(k)$ bins so any that we find in the bootstrap realizations are
simply noise.  Visually comparing the error matrices from the mock
spectra and the real data we find that they look very similar.  There
is no obvious evidence that the errors in the real data are more 
correlated than the theoretically uncorrelated mock data (both become
strongly correlated at very high $k$ where the window function 
becomes important).  We will simply use the diagonal elements of the
bootstrap error matrix and treat the bin errors as independent.
We correct the errors on each bin separately by comparing the 
true and bootstrap errors in the mock spectra.  These corrections
are only significant ($\sim 30$\%) on the longest wavelength modes.
Note that using independent bootstrap errors on each bin still takes
into account the correlation between the individual $k$ modes that
are averaged to give the bin value.  The measurement of the amplitude
shows that at the redshifts
in question the modes with $k<0.04~(\kms)^{-1}$ should not be fully
non-linear supporting the supposition that the correlations between 
bins is not large.

\section{THE \lya FOREST FLUX POWER SPECTRUM IN THE LARGE SCALE LIMIT}

  This Appendix shows that the flux power spectrum in the \lya
forest must indeed be proportional to the mass power spectrum in the
limit of large scales, as was first assumed by citet{cwk98}.
We shall first consider the simple case where the effect of peculiar
velocities is neglected.

  Let $P_0(F)$ be the flux probability distribution, and
$P(F|\bar\delta)$ be the conditional distribution of $F$ given that the
overdensity smoothed over a certain scale $R_s$ around the point where
the flux $F$ is observed is $\bar\delta$. We
consider the correlation of the transmitted flux at two points, $F_1$
and $F_2$, separated by a distance much larger than $R_s$, but where
$R_s$ is still a sufficiently large scale for linear theory to be valid.
In this case, the correlation of $F_1$ and $F_2$ should be exclusively
due to the correlation of $\bar\delta_1$ and $\bar\delta_2$, the
smoothed densities around the points 1 and 2. In other words, the joint
probability distribution $P_2(F_1,F_2)$ should be equal to the product
$P(F_1|\bar\delta_1) P(F_2|\bar\delta_2)$ (notice that the correlation
of the whole deformation tensor at the two points should enter here at
the same order as the density correlation; we are assuming that the
constrained probability distribution of the flux depends mostly on the
overdensity, and not the other components of the deformation tensor).
Since $\bar\delta \ll 1$, we can write $P(F|\bar\delta) = P_0(F) +
\bar\delta \, P_{\delta}(F)$, where 
$P_\delta(F) = [dP(F|\bar\delta)/d\bar\delta]_{\bar\delta=0}$.
The flux correlation function is then given by:

\begin{eqnarray}
  < F_1 F_2 > &=& \int dF_1\, dF_2 \,
< \left[ P_0(F_1) + \bar\delta_1 P_{\delta}(F_1) \right] \,
\left[ P_0(F_2) + \bar\delta_2 P_{\delta}(F_2) \right] ~ F_1\, F_2 > 
\nonumber \\  &=&
 < \bar\delta_1 \bar\delta_2 > \int dF_1\, dF_2 \,
P_{\delta}(F_1) P_{\delta}(F_2) ~ F_1\, F_2 \nonumber \\  &=&
  \xi(x_{12}) \, \cdot \, \left[ \int dF\, P_{\delta}(F)\, F \right]^2  
\nonumber \\  &=&
\xi(x_{12}) \, \cdot \, \left( { d\bar F \over d\bar\delta } \right)^2 ~,
\end{eqnarray}
where $\xi(x_{12}) = < \bar\delta_1 \bar\delta_2 >$ is the density correlation
function at the separation $x_{12}$ between points 1 and 2, and
$\bar F(\bar\delta) = \int dF\, F\, P(F,\bar\delta)$. This gives
an expression for the $B$ factor defined in \S 5.7.

  When peculiar velocities are included, there should be an additional
dependence of $P(F)$ on the peculiar velocity gradient along the line
of sight, $\eta = H^{-1} dv_p/dl$ (where $d/dl$ is the derivative along
the line of sight), averaged over the same scale $R_s$. We can linearize
as before $P(F,\bar\delta,\bar\eta) = P_0(F) + \bar\delta P_{\delta}(F)
+ \bar\eta P_{\eta}(F)$, and define the factor
\begin{equation}
  b_F = \left(\int dF\, P_{\delta}(F)\, F \right) / 
\left(\int dF\, P_{\eta}(F)\, F \right) ~.
\end{equation}
The quantity $b_F$ is a different ``bias factor'' for the \lya forest,
this one intended for peculiar velocity calculations. Defining the new
variable $\bar\delta_z = \bar\delta + b_F^{-1} \bar\eta$, we obtain
\begin{equation}
  < F_1 F_2 > = \xi_z(x_{12}) \, \cdot \,
\left( { \partial \bar F \over \partial \bar\delta } \right)^2 ~,
\end{equation}
where $\xi_z$ is the redshift space correlation function along the line
of sight of $\delta_z$, which has bias factor $b_F$. Thus, we see that
both $b_F$ and $B = \partial \bF / \partial \bar{\delta}$ are functions
of the flux distribution $P_0(F)$, and of how this distribution varies
with the overdensity and the peculiar velocity gradient smoothed over
a large scale around the point where the flux is being measured. This
implies a complex dependence of $b_F$ and $B$ on all the variables
determining the \lya forest properties, such as the fluctuation
amplitude on the Jeans scale, the mean transmitted flux, and the
temperature-density relation.


\begin{thebibliography}{}
\bibitem[Abel \& Mo(1998)]{am98} Abel, T. \& Mo, H. J. 1998, ApJ, 
             494, L151
\bibitem[Arons(1972)]{a72} Arons, J. 1972, ApJ, 172, 553	   
\bibitem[Bahcall \& Salpeter(1965)]{bs65} Bahcall, J. N. \& Salpeter, 
             E. E. 1965, \apj, 142, 1677  	     
\bibitem[Bardeen \etal(1986)]{bbk86} Bardeen, J. M., Bond, J. R., Kaiser, N., \& Szalay, A.
             S. 1986, ApJ, 304, 15 
\bibitem[Barlow \& Sargent(1997)]{bs97} Barlow, T. A., \& Sargent, W. L. W. 1997, AJ, 113, 136

\bibitem[Bechtold \etal(1994)]{b94} Bechtold, J., Crotts, A. P. S.,
Duncan, R. C., \& Fang, Y. 1994, ApJ, 437, L83

\bibitem[Bi(1993)]{b93} Bi, H. 1993, ApJ, 405, 479

\bibitem[Bond \etal(1988)Bond, Szalay, \& Silk]{bss88} Bond, J. R., Szalay, A. S., \& Silk, J. 1988, ApJ, 324,
             627
\bibitem[Bryan \& Machacek(1999)]{bm99} Bryan, G. L., \& Machacek, M. E. 1999, ApJ, submitted
              (astro-ph/9906459)
\bibitem[Burles \& Tytler(1998)]{bt98} Burles, S. \& Tytler, D. 1998, ApJ, 499, 699	      
\bibitem[Cen \etal(1994)]{cmo94} Cen, R., Miralda-Escud\'e, J., Ostriker, J. P., \& Rauch,
             M. 1994, ApJ, 437, L9
\bibitem[Croft \etal(1999)Croft, Hu, \& Dav\'e]{chd99} Croft, R. A. C., 
             Hu, W., \& Dav\'e, R. 1999, Phys. Rev. Lett., 83, 1092 
\bibitem[Croft \etal(1998)]{cwk98} Croft, R. A. C., Weinberg, D. H., Katz, N., \&
             Hernquist, L. 1998, ApJ, 495, 44 
\bibitem[Croft \etal(1999)]{cwp99} Croft, R. A. C., Weinberg, D. H., Pettini, M., 
             Hernquist, L., \& Katz, N. 1999, ApJ, 520, 1	     
\bibitem[Crotts \& Fang(1998)]{cf98} Crotts, A. P. S., \& Fang, Y. 1998, ApJ, 502, 16

\bibitem[Dinshaw \etal(1994)]{d94} Dinshaw, N., Impey, C. D., Foltz, C. B.,
Weymann, R. J., \& Chaffee, F. H. 1994, ApJ, 437, L87

\bibitem[Gnedin \& Hui(1998)]{gh98} Gnedin, N. Y., \& Hui, L. 1998,
MNRAS, 296, 44

\bibitem[Hernquist \etal(1996)]{hkw96} Hernquist, L., Katz, N.,
Weinberg, D. H., \& Miralda-Escud\'e, J. 1996, ApJ, 457, L51   

\bibitem[Hui \& Gnedin(1997)]{hg97} Hui, L., \& Gnedin, N. Y. 1997,
MNRAS, 292, 27

\bibitem[Hui(1999)]{h99} Hui, L. 1999, ApJ, 516, 519
\bibitem[Jenkins \& Ostriker(1991)]{jo91} Jenkins, E. B., \& Ostriker,
J. P. 1991, ApJ, 376, 33

\bibitem[Ma(1996)]{m96} Ma, C. 1996, ApJ, 471, 13

\bibitem[McDonald \& Miralda-Escud\'e(1999a)]{mm99a} McDonald, P., \&
Miralda-Escud\'e, J. 1999a, ApJ, 518, 24

\bibitem[McDonald \& Miralda-Escud\'e(1999b)]{mm99b} McDonald, P., \&
Miralda-Escud\'e, J. 1999b, ApJ, 519, 486

\bibitem[McGill(1990)]{m90} McGill, C. 1990, MNRAS, 242, 544

\bibitem[Miralda-Escud\'e \etal(1996)]{mco96} Miralda-Escud\'e, J., 
             Cen, R., Ostriker, J. P., \& Rauch, M. 1996, ApJ, 471, 582
\bibitem[Miralda-Escud\'e \& Rees(1993)]{mr93} Miralda-Escud\'e, J., \& Rees, M. J. 1993, MNRAS, 260, 617 
\bibitem[Miralda-Escud\'e \& Rees(1994)]{mr94} Miralda-Escud\'e, J., \& Rees, M. J. 1994, MNRAS, 266, 343
\bibitem[Nusser \& Haehnelt(1999)]{nh99} Nusser, A., \& Haehnelt, M. 1999, MNRAS, 303, 179	     
\bibitem[Press \etal(1992)]{ptv92} Press, W., Teukolsky, S., Vetterling, W., \& Flannery,
             B. 1992, Numerical Recipes in C
             (2d ed.; Cambridge:Cambridge University Press) 
\bibitem[Press, Rybicki, \& Schneider(1993)]{prs93} Press, W. H.,
         Rybicki, G. B., \& Schneider, D. P. 1993, ApJ, 414, 64
\bibitem[Prochaska \& Wolfe(1997)]{pw97} Prochaska, J. X., \& Wolfe, A. M. 1997, ApJ, 487, 73
\bibitem[Prochaska \& Wolfe(1998)]{pw98} Prochaska, J. X., \& Wolfe, A. M. 1998, ApJ, 507, 113
\bibitem[Rauch \etal(1997)]{rms97} Rauch, M., Miralda-Escud\'e, J., Sargent, W. L. W., 
             Barlow, T. A., Weinberg, D. H., Hernquist, L., Katz, 
             N., Cen, R., \& Ostriker, J. P. 1997, ApJ, 489, 7
\bibitem[Rauch(1998)]{r98} Rauch, M. 1998, ARA\&A, 36, 267
\bibitem[Rauch \etal(1992)]{rcc92} Rauch, M., Carswell, R. F., 
             Chaffee, F. H., Foltz, C. B., Webb, J. K., Weymann, R. J.,
	     Bechtold, J., \& Green, R. F. 1992, ApJ, 390, 387
\bibitem[Rees(1986)]{r86} Rees, M. J. 1986, MNRAS, 218, 25P

\bibitem[Ricotti \etal(1999)
         Ricotti, Gnedin, \& Shull]{rgs99} Ricotti, M., Gnedin, N. Y., \&
         Shull, J. M. 1999, ApJ, submitted (astro-ph/9906413)

\bibitem[Schaye \etal(1999)]{stl99} Schaye, J., Theuns, T., Leonard, A.,
Efstathiou, G. 1999, MNRAS, in press (astro-ph/9906271)

\bibitem[Scherrer \& Weinberg(1998)]{sw98} Scherrer, R. J., \&
             Weinberg, D. H. 1998, ApJ, 504, 607

\bibitem[Smette \etal(1992)]{s92} Smette, A., Surdej, J., Shaver, P. A.,
         Foltz, C. B., Chaffee, F. H., Weymann, R. J., Williams, R. E.,
         \& Magain, P. 1992, ApJ, 389, 39

\bibitem[Smette \etal(1995)]{s95} Smette, A., Robertson, J. G., Shaver, P. A.,
         Reimers, D., Wisotzki, L., \& K\"ohler, Th. 1995, A\&AS, 113, 199

\bibitem[Theuns \etal(1998)]{tle98} Theuns, T., Leonard, A., Efstathiou, G.,
         Pearce, F. R., \& Thomas, P. A. 1998, MNRAS, 301, 478

\bibitem[Vogt \etal(1994)]{vab94} Vogt, S. S., \etal 1994, SPIE 2198, 362

\bibitem[Wadsley \& Bond(1996)]{wb96} Wadsley, J. W., \& Bond, J. R. 1996,
         preprint, (astro-ph/9612148)     

\bibitem[Wang \etal(1999)]{wcos99} Wang, L., Caldwell, R. R., Ostriker, J. P.,
        \& Steinhardt, P. J. 1999, ApJ, in press (astro-ph/9901388)

\bibitem[Weinberg \etal(1997)]{wmh97} Weinberg, D. H., Miralda-Escud\'e, J.,
         Hernquist, L., \& Katz, N. 1997, ApJ, 490, 564 

\bibitem[Weinberg \etal(1999)]{wch99} Weinberg, D. H., Croft, R. A. C.,
         Hernquist, L., Katz, N., \& Pettini, M. 1999, ApJ, 522, 563

\bibitem[Zhang \etal(1995)Zhang, Anninos, \& Norman]{zan95} Zhang, Y.,
         Anninos, P., \& Norman, M. L. 1995, ApJ, 453, L57 

\bibitem[Zhang \etal(1997)]{zan97} Zhang, Y., Anninos, P., Norman, M. L., \&
         Meiksin, A. 1997, ApJ, 485, 496 		     	   

\end{thebibliography}
\end{document}